\documentclass[review,onefignum,onetabnum]{article}

\usepackage{latexsym}
\usepackage{amssymb,amsbsy,amsmath,amsfonts,amssymb,amscd,amsthm}
\usepackage{graphicx}
\usepackage{hyperref}
\usepackage{mathtools}
\usepackage{pdfsync}
\usepackage{epsfig,epstopdf}
\usepackage{xcolor}
\usepackage{hyperref}
\usepackage[title]{appendix}
\usepackage{verbatim}
\usepackage{lipsum}
\usepackage{float}
\usepackage[font=small]{caption}
\usepackage{subcaption}
\usepackage[makeroom]{cancel}
\usepackage{booktabs}
\usepackage{tabularx}
\usepackage{braket}
\usepackage{comment}

\DeclarePairedDelimiter {\abs}{\lvert}{\rvert}
\newcolumntype{Y}{>{\raggedright\arraybackslash}p{5cm}}
\newcolumntype{Z}{>{\raggedright\arraybackslash}p{1.25cm}}

\newcommand{\R}{\mathbb{R}}

\newcommand{\di}{\,\mathrm{d}}
\newcommand{\pt}{\partial_t}
\newcommand{\px}{\partial_x}
\newcommand{\py}{\partial_y}

\newcommand{\pyy}{\partial_{yy}^2}

\newcommand{\expzero}{case 1}
\newcommand{\exponea}{case 2a}
\newcommand{\exponeb}{case 2b}
\newcommand{\exptwoa}{case 3a}
\newcommand{\exptwob}{case 3b}
\newcommand{\david}[1]{{{#1}}}
\newcommand{\giulia}[1]{{{#1}}}

\newcommand{\beq}{\begin{equation}}
\newcommand{\eeq}{\end{equation}}
\newcommand{\beqa}{\begin{eqnarray}}
\newcommand{\eeqa}{\end{eqnarray}}

\DeclareMathOperator{\dive}{div}
\DeclareMathOperator{\supp}{supp}
\DeclareMathOperator{\diam}{diam}

\providecommand{\keywords}[1]
{
	\small	
	\textbf{\textit{Keywords---}} #1
}

\renewcommand{\vec}{\boldsymbol}
\renewcommand{\epsilon}{\varepsilon}

\numberwithin{equation}{section}
\numberwithin{prpstn}{section}
\numberwithin{ass}{section}
\numberwithin{rmrk}{section}

\title{A phenotype-structured mathematical model for the influence of hypoxia on oncolytic virotherapy\thanks{Corresponding author: David Morselli (d.morselli@ucl.ac.uk)\\
This research was partially supported by the Italian Ministry of Education, University and Research (MIUR) through the “Dipartimenti di Eccellenza” Programme (2018-2022) – Dipartimento di Scienze Matematiche “G. L. Lagrange”, Politecnico di Torino (CUP: E11G18000350001). DM, GC and MED are members of GNFM (Gruppo Nazionale per la Fisica Matematica) of INdAM (Istituto Nazionale di Alta Matematica). We also acknowledge the support of the Australian National Health and Medical Research Council, through grant NHMRC IDEAS 2013058. Part of this work was performed on the OzSTAR national facility at Swinburne University of Technology. The OzSTAR program receives funding in part from the Astronomy National Collaborative Research Infrastructure Strategy (NCRIS) allocation provided by the Australian Government, and from the Victorian Higher Education State Investment Fund (VHESIF) provided by the Victorian Government.}
}
\author{David Morselli,\thanks{Department of Mathematics, University College London, 25 Gordon Street, London WC1H 0AY, United Kingdom} \thanks{Department of Mathematical Sciences ``G. L. Lagrange'', Politecnico di Torino, Corso Duca degli Abruzzi 24, 10129 Torino, Italy} \thanks{Department of Mathematics, School of Science, Computing and Engineering Technologies, Swinburne University of Technology, John St, 3122, Hawthorn, VIC, Australia} \thanks{Department of Mathematics ``G. Peano'', Università di Torino, Via Carlo Alberto 10, 10124 Torino, Italy}
	\and
	Giulia Chiari,\thanks{BCAM – Basque Center for Applied Mathematics, Mazarredo Zumarkalea, 14, Abando, 48009 Bilbo, Spain} \footnotemark[3] \footnotemark[4] \footnotemark[5]
	\and
	Federico Frascoli,\footnotemark[4] 
	\and
	Marcello Edoardo Delitala\footnotemark[3]
}

\begin{document}

\maketitle

\begin{abstract}
The effectiveness of oncolytic virotherapy is significantly affected by several elements of the tumour microenvironment, which reduce the ability of the virus to infect cancer cells. In this work, we focus on the influence of hypoxia on this therapy and develop a novel continuous mathematical model that considers both the spatial and epigenetic heterogeneity of the tumour. We investigate how oxygen gradients within tumours affect the spatial distribution and replication of both the tumour and oncolytic viruses, focusing on regions of severe hypoxia versus normoxic areas. Additionally, we analyse the evolutionary dynamics of tumour cells under hypoxic conditions and their influence on susceptibility to viral infection. Our findings show that the reduced metabolic activity of hypoxic cells may significantly impact the virotherapy effectiveness; the knowledge of the tumour's oxygenation could, therefore, suggest the most suitable type of virus to optimise the outcome. The combination of numerical simulations and theoretical results for the model equilibrium values allows us to elucidate the complex interplay between viruses, tumour evolution and oxygen dynamics, ultimately contributing to developing more effective and personalised cancer treatments.
\end{abstract}


\section{Introduction}

Among cancer therapies, oncolytic virotherapy stands out as a promising avenue that harnesses the natural capabilities of viruses to selectively target and destroy cancer cells while sparing healthy tissues, \cite{blanchette23,fountzilas17,Kelly2007651review,lawler17,russell18}. Despite its potential, the clinical efficacy of oncolytic virotherapy faces significant challenges, many of which stem from the complex dynamics of the tumour microenvironment (TME) \cite{jin21}.

The TME significantly influences viral distribution and therapeutic efficacy. Factors such as extracellular matrix composition, immune cell infiltration and hypoxic regions can impede viral penetration, replication and spread within the tumour \cite{wojton10}. These barriers create unpredictable, stochastic events that affect the consistency of viral delivery and the overall success of the therapy \cite{jin21}. Another interesting interaction involves the neovasculature that the tumour originates through angiogenesis: most oncolytic viruses disrupt it by targeting tumour-associated vascular endothelial cells \cite{jin21,wojton10}. On the one hand, the decrease in nutrient inflow contributes to slowing down cancer growth and, in this respect, oncolytic virotherapy could act in the same way as antiangiogenic therapy \cite{santry20}. On the other hand, blood vessels also play a vital role in the arrival of viral particles and immune cells \cite{santry20}.

The inadequate vascularisation within tumours reduces oxygen availability: this hypoxic condition is a hallmark of solid tumours, significantly affects the tumour's evolution and profoundly impacts treatment efficacy \cite{zhuang23}. Hypoxic regions within tumours promote aggressiveness by harbouring cells with reduced metabolic activity and heightened resistance to therapies  \cite{zhuang23}. In the context of oncolytic virotherapy, the impact of hypoxia on therapeutic efficacy is not straightforward \cite{shengguo11}. In general, hypoxic regions within tumours harbour cells less susceptible to viral infection and replication due to reduced metabolic activity and altered cellular signalling pathways; furthermore, the physiological adaptations of tumour cells to hypoxia, such as enhanced glycolytic metabolism and resistance to apoptosis, can confer resistance to viral-induced cell death. It is important to remark that the previous considerations are not universal: indeed, some particular oncolytic viruses can specifically target receptors that are upregulated in case of the lack of oxygen \cite{shengguo11,sadri23}.

Understanding how hypoxia influences the interaction between OVs and tumour cells is crucial for optimising treatment strategies and overcoming therapeutic resistance. Mathematical modelling is pivotal in unravelling these complexities and optimising therapeutic outcomes \cite{byrne10}. The dynamics of viral replication and tumour growth have been analysed through several modelling approaches, ranging from ordinary differential equations (ODEs) \cite{almuallem21,eftimie18,jenner19,jenner18insights,komarova10,novozhilov06,storey20,vithanage23,wodarz01} and partial differential equations (PDEs) \cite{alzahrani19,friedman03,friedman18,kim14,pooladvand22,wu01,wu04} to stochastic agent-based models \cite{jenner20voronoi,storey21,surendran23,wodarz12} and hybrid multi-scale models \cite{jenner22,paiva09}. Some of these models also take into account aspects related to the tumour microenvironment, such as spatial constraints due to the extracellular matrix \cite{alzahrani19,jenner22,kim14,pooladvand22} and interactions with the immune system \cite{almuallem21,eftimie18,storey20,vithanage23,friedman18,jenner22,wodarz01,wu04}. Only a few mathematical models describe virotherapy in hypoxic conditions. In Ref. \cite{ramaj23}, the authors consider ODEs and nested ODEs (corresponding to infections in adjacent lymph nodes) under the assumption that oxygen concentration directly influences the infection rate. The only spatial model that we know in this context is the one presented in Ref. \cite{boemo19}, in which the difficulties of treating hypoxic regions with standard therapies motivate the use of macrophages that release oncolytic viral particles when experiencing low oxygen concentrations. The lack of spatial models that consider the influence of oxygen concentration on virotherapy effectiveness motivates the present work.

Even in the absence of viral infection, non-genetic changes in cancer cells significantly affect the metabolic activity and these evolutionary dynamics can be analysed with mathematical models \cite{bosque23}. In particular, the influence of hypoxia on tumour dynamics has been investigated through several approaches, using either discrete compartments that evolve according to different dynamics \cite{astanin08,Gatenby,martinez12,rocha2021persistent} or a continuous spectrum of adaptation levels \cite{ardaseva2020fluctuating,ardaseva20periodic,gallaher}. The latter approach in the continuous settings leads to integro-differential equations (IDEs) and, when spatial heterogeneity is included, to partial integro-differential equations (PIDEs) \cite{chiari23heterogeneity,fiandaca2020mathematical,lorenzi2018role,villa21modeling}. We build upon this modelling approach and consider the heterogeneous effects of oncolytic virotherapy within the framework of spatial and phenotypic variability.

A common characteristic of continuous structured models is 
the presence of a \textit{trade-off} between different features \cite{lorenzi24preprint}. At a cellular level, the term trade-off refers to the presence of mutually exclusive properties, usually caused by the need for metabolic optimisation, with the possibility that an improvement
in one aspect necessarily leads to a deterioration in another. One of the most considered trade-offs in the context of hypoxia involves the proliferation rate and resistance to hypoxic conditions. This can be interpreted with the following meaning: cells whose proliferative performance is significantly enhanced by oxygen are also more easily induced to die in harsh conditions. In contrast, cells that are not very sensitive to increased proliferation due to the abundance of nutrients maintain a state of quiescence or low proliferation in the event of a lack of oxygen, without incurring cell death.

In the context of oncolytic virotherapy, different oncolytic virus types perform effectively or poorly in hypoxic condition according to their specific features, as described above. This, together with the previous trade-off, leads to two opposite scenarios. Viruses that are more effective in normoxic conditions lead to greater infectivity in highly proliferating tumours. In contrast, more efficient viruses under hypoxic conditions will lead to greater infectivity in tumours highly resistant to hypoxia. The two modelling approaches resulting from these scenarios are represented in Fig. \ref{fig:tradeoffs}.

More in detail, our model incorporates both spatial and phenotypic heterogeneity of tumour cells, along with the dynamics of oxygen concentration and viral infection; the model is formulated using a combination of partial differential equations (PDEs) and partial integro-differential equations (PIDEs), explicitly accounting for the spatial gradients of oxygen within the tumour. Such a modelling approach allows us to perform a formal asymptotic analysis of simplified settings to compute the homogeneous equilibrium values. 

It is important to remark that the use of a continuous trait variable to describe the system's heterogeneity has been widely employed for equations that share some formal similarities to the ones analysed in the present work, both in the epidemiological settings \cite{almeida21,bernardi22,lorenzi24,lorenzi21} and the ecological settings \cite{dearaujo24,delitala13evolution}. In oncolytic virotherapy, structured populations have been used in Ref. \cite{karev06} to model various kinds of heterogeneity, including susceptibility to infections, death rates, and virulence; neither of them considers trade-offs similar to the ones we aim to investigate.
A slightly different approach involves the use of an age structure: indeed, the inclusion of the time from the infection as a structuring variable for the infected individuals dates back to the first modern SI model \cite{kermack27}. Analogous models were also adopted in Refs. \cite{ding22,gao22} in relation to the time from infection for oncolytic virotherapy and in Ref. \cite{portillo24} in relation to the temporal evolution of stem cells. 

As previously mentioned, in this work, we build upon the model presented in Ref. \cite{lorenzi21} and include spatial heterogeneity and environmental factors that influence the dynamics (as in Refs. \cite{chiari23heterogeneity,chiari23radio}); to our knowledge, both aspects constitute a novelty with respect to the existing literature. 

\giulia{It is important to note that the choice to integrate both epigenetic and spatial characterisation, in addition to being novel, opens new avenues for investigating the epigenetic mapping of tumours. This approach aligns with the current biological and oncological research direction that increasingly leverages spatial transcriptomics. Within the framework of dose painting, which characterises the study of other widespread therapies such as radiotherapy, this strategy paves the way for optimisation studies, both in terms of the spatial characterisation of oncolytic virus delivery in situ and of the combination of such strategies with more spatially specific treatments. These could, in turn, target regions where, due to local environmental features, this therapy may otherwise prove less effective.}

In this work, we aim to

\begin{itemize}
	\item[-] characterise the impact of hypoxia on viral infection and investigate how oxygen gradients within tumours affect the spatial distribution and replication of the tumour and the oncolytic viruses, with a focus on regions of severe hypoxia versus normoxic areas;
	\item[-] explore evolutionary dynamics and analyse how hypoxia-induced adaptations in tumour cells influence their susceptibility to viral infection, the epigenetic composition of the tumour, and the emergence of resistant phenotypes over time;
	\item[-] consider the trade-off between proliferation rate and resistance to hypoxia in view of optimising therapeutic strategies to enhance the possibility to exploit different oncolytic virus types and corresponding features to enhance treatment efficacy.
\end{itemize}

The rest of the paper is organised as follows: Section  \ref{sec:model_hyp} provides the mathematical formulation of the model, detailing the integration of viral dynamics, tumour evolution, and spatial oxygen gradients, and a brief theoretical analysis. In Section \ref{sec:results}, we present numerical simulations and explore the impact of hypoxia on the efficacy of oncolytic virotherapy; we also mention the situation of a virus that specifically targets hypoxic cells, looking towards the combination of several therapies. Section \ref{sec:conclusion} concludes the paper by summarising the key findings and discussing the implications of these findings for optimising treatment strategies.

\section{Modelling framework}
\label{sec:model_hyp}
We consider the epigenetic heterogeneity of uninfected cancer cells and assume that it affects resistance to hypoxia in addition to proliferation and infection. The dynamics of infected cells, instead, are not affected by epigenetic characteristics, hence we model them as a homogeneous population. We first describe the model and then carry out a simple asymptotic analysis to characterise the equilibrium of the problem under the assumption of stationary oxygen.

\subsection{Model description}
Let us denote by $t\in [0,+\infty)$ the time variable, by $\vec{x}\in \Omega$ the space variable, with $\Omega\subset\R^2$ and by $y\in Y$ the epigenetic variable, with $Y\coloneqq [0,1]$.
The variable $y$ is intended as the normalized joint expression of a set of genes selected as responsible for determining the potential features of the cell. In this case, we consider the three characteristics involved in the trade-offs (proliferation rate, resistance to hypoxic condition and infectability).

We consider uninfected and infected cancer cells, whose densities are described respectively by the functions $u\colon [0,+\infty)\times\Omega\times Y \to [0,+\infty)$ and $I\colon [0,+\infty)\times\Omega\to [0,+\infty)$. We also define the uninfected total cell density as
\begin{equation}
	\label{eq_U}
	U(t,\vec{x})\coloneqq \int_Y u(t,\vec{x},y) \di y
\end{equation}
and the total cancer cell density as
\begin{equation}
	\rho(t,\vec{x}) \coloneqq I(t,\vec{x}) + U(t,\vec{x}).
	\label{eq_rho}
\end{equation}
Finally, we consider viral density, described by the function $v\colon [0,+\infty)\times\Omega\to [0,+\infty)$, and oxygen concentration, described by the function $O\colon [0,+\infty)\times\Omega\to [0,+\infty)$. We now describe in detail the rules governing all the dynamics.

\paragraph{Uninfected cancer cells}
Uninfected cells may move via pressure-driven movement, change their epigenetic trait, reproduce, become infected and die due to environment-driven selective pressure. We assume a trade-off between proliferation and resistance to hypoxia. In this sense, we consider $y$ as the level of expression of a set of genes responsible for this trade-off and normalise it so that $y=0$ and $y=1$ are, respectively, the lowest and highest possible expressions: $y=0$ corresponds to highest intrinsic proliferation rate and lowest resistance to hypoxia; conversely  $y=1$ corresponds to lowest intrinsic proliferation rate and highest resistance to hypoxia. We also assume that the resistance to the oncolytic viral infection is affected by the same set of genes, with the exact dependence determined by the kind of virus under investigation. The evolution of uninfected cells is described by the equation
\begin{align*}
	\pt u(t,\vec{x},y)=& R(y,\rho(t,\vec{x}),O(t,\vec{x}),v(t,\vec{x}))\, u(t,\vec{x},y)\\
	&+ D_y \underbrace{\pyy u(t,\vec{x},y)}_{\text{random mutation}} +\underbrace{ D_{\vec{x}} \dive_{\vec{x}} (u(t,\vec{x},y) \nabla\rho(t,\vec{x}))}_{\text{pressure-driven movement}}
\end{align*}
with
\[
R(y,\rho,O,v)=\big(\!\!\!\!\underbrace{P(y,\rho)}_{\text{proliferation}}-\underbrace{S(y,O)}_{\text{sel. pressure}}\big) -\underbrace{\beta(y)\; v}_{\text{infection}}.
\]
Random epigenetic mutations are described by a diffusive term in $y$ with coefficient $D_{y}$. Cancer cells also move with coefficient $D_{\vec{x}}$ in space against the gradient of the total cancer cell density $\rho(t,\vec{x})$. Furthermore, uninfected cancer cells proliferate at a rate determined by the intrinsic proliferation rate $p(y)$ and the local cancer cell density $\rho(t,\vec{x})$, according to the logistic growth term
\begin{equation}
	P(y,\rho) = p(y) \left(1-\dfrac{\rho}{K}\right).
	\label{eq_P}
\end{equation}
Uninfected cells may also die because of the environment-driven causes, here determined by oxygen concentration.
In the model we refer to this term as the ``selective pressure'' in the sense that, being the death rate dependent of the distance of the epigenetic state of the cell from an optimal one, it tends to benefit this state in evolutionary terms. The biological assumption at the base of this formulation is that each oxygen level determines an epigenetic state which can be considered ``optimal'' in terms of resistance under that oxygenation condition. Note that here we are not considering the fittest trait resulting from the overlap of all the dynamics included in the model, but only the epigenetic state that ensures a lower mortality rate induced by environmental conditions. The such defined optimal trait is determined by the function 
\begin{equation}
	\varphi(O)\coloneqq 
	\begin{cases}
		1 \quad&\text{if } O\leq O_m \\[8pt]
		\dfrac{O_M-O}{O_M-O_m} \quad&\text{if } O_m<O<O_M \\[8pt]
		0 \quad&\text{if } O\geq O_M 
	\end{cases}
	\label{eq_phi}
\end{equation}
where $O_M$ and $O_m$ denote respectively the normoxic and hypoxic thresholds. We assume that the optimal trait is $y=0$ in normoxic situations (i.e, when the oxygen concentration is above $O_M$) and $y=1$ in severely hypoxic situations (i.e., when the oxygen concentration is below $O_m$), coherently with our interpretation of the epigenetic variable $y$; for intermediate oxygen concentrations, the function $\varphi$ provides a linear interpolation between these two values.
Furthermore, the selective pressure is expressed through a quadratic function of the distance of trait $y$ from the optimal one:
\begin{equation}
	S(y,O) =  \eta (y-\varphi(O))^2
	\label{eq_S}
\end{equation}	
where $\eta$ defines the time scale at which the process takes place. Finally, the virus infects uninfected cancer cells according to the viral density and the infection rate $\beta(y)$.

\begin{figure}
	\centering
	\includegraphics[width=0.75\linewidth]{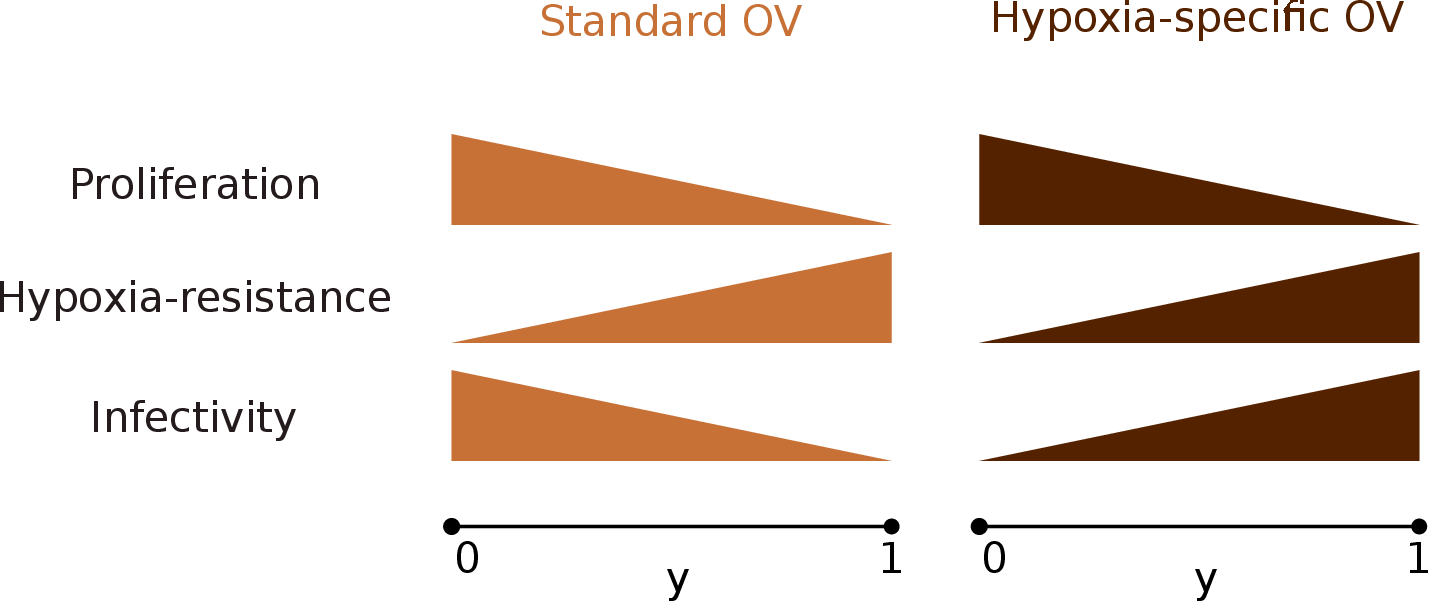}
	\caption{Graphical representation of trade-offs involving proliferation rate (row 1), hypoxia-resistance features (row 2) and infectivity potential rate (row 3) in the case of standard (left) and hypoxia-specific (right) viruses. The correspondence with the values assumed by epigenetic variable $y\in [0,1]$ are indicated. These represent the cell potentiality, which effective effect on the cell phenotype is determined according to the environmental condition and the evolutionary scenarios.}
	\label{fig:tradeoffs}
\end{figure}

Note that $y$ is directly involved in the environment-driven selective pressure term; it also affects proliferation and infection through the coefficients $p(y)$ and $\beta(y)$. To catch the above-described trade-offs, we set
\begin{equation}
	p(y) = p_M - (p_M-p_m) y, \qquad \beta(y) = \beta_M +  (\beta_m-\beta_M) y
	\label{eq_pbeta}
\end{equation}
being $p_M$ and $\beta_M$ the values associated with cells with the minimal epigenetic trait $y=0$ and $p_m$ and $\beta_m$ the values that characterised cells with the maximal epigenetic trait $y=1$. Observe that we assume $p_M>p_m$ and so the function $p$ is decreasing in $y$, coherently with our initial assumptions. On the other hand, the behaviour of the function $\beta$ depends on the specific biological situation under investigation: the assumption $\beta_M>\beta_m$ leads to a function $\beta$ decreasing in $y$ and models the decrease of viral infectivity associated to hypoxic cells; the choice $\beta_m>\beta_M$ leads instead to an increasing $\beta$ and corresponds to a virus that specifically targets hypoxic cells.
Since from a therapeutical point of view it is known that most treatment strategies improve their efficiency on proliferative cells in well-oxygenated environments, we refer to viruses behaving as in the first case as a ``standard oncolytic viruses'', and to the second one as an ``hypoxia-specific oncolytic viruses''. These two scenarios are represented in Figure \ref{fig:tradeoffs}.

Note that the effect of oxygenation on the infectivity of oncolityc viruses opens the way to two different modelling choices: one expressed with a direct dependence on oxygen, the other with a dependence on the epigenetic variable $y$ involved in determining the rate of resistance to hypoxia of the cell. Although both choices are considered equally valid, we adopt the second modeling strategy in this context. Our aim is to trace how epigenetic variations in tumour cells, implemented as a resistance mechanism in hypoxic conditions, affect different levels of infectious efficacy, rather than attributing these differences directly to oxygen deprivation altering the virus’s capabilities.

We remark that the terms for proliferation and selective pressure follow the modelling choices adopted in Ref. \cite{chiari23heterogeneity}. In absence of infections and epigenetic diffusion, the cell density always grows up to the same carrying capacity $K$ irrespectively of the oxygen level: indeed, $u(t,\vec{x},y)=K\delta_{\varphi(O(\vec{x}))}(y)$ is trivially a steady state (see also Eq. \eqref{eq:eq_K} in this respect). This constitutes a significant difference with respect to some previous modelling approaches of hypoxic tumours (such as the ones employed in Refs. \cite{ardaseva2020fluctuating,ardaseva20comparative,fiandaca2020mathematical,villa21modeling}) and allows us to consider virotherapy as the only cause of cancer reduction in hypoxic conditions. In presence of epigenetic diffusion, it is reasonable to expect a smaller equilibrium value of $\rho$ due to the coexistence of subpopulations with different epigenetic characteristics and the resulting action of the term $S$.

\paragraph{Infected cancer cell} 
Infected cells may move and die; uninfected cells may also become infected. The dynamics are described by the equation
\begin{equation*}
	\pt I(t,\vec{x})= \underbrace{D_{\vec{x}} \dive (I(t,\vec{x})\nabla \rho(t,\vec{x}))}_{\text{pressure-driven movement}} + \underbrace{v(t,\vec{x}) \int_Y \beta(y) u(t,\vec{x},y) \di y}_{\text{infection}}-\underbrace{q_I I(t,\vec{x})}_{\text{death}}.
\end{equation*}
The movement follows the same law as that of uninfected cells, as there is no reason to assume that the infection has some effect on that. All the susceptible cells that undergo infection are collected in the same population. Finally, infected cells die at rate $q_I$. This equation does not take into account the uptake of free viruses by cancer cells: although a few models include this process \cite{bajzer08,pooladvand21}, it is usually assumed to be negligible \cite{almuallem21,baabdulla23,jenner19,jenner18insights,storey20,vithanage23}.

We assume that infected cells do not proliferate, as the virus disrupts the cellular machinery, and are not affected by environmental conditions due to their short life. Consequently, they lack all the processes the epigenetic trait affects, which motivates their absence of epigenetic structure. This approach is the same one adopted in Ref. \cite{lorenzi21}, with the only differences that we here consider a spatial structure and infections mediated by a viral population.

\paragraph{Oncolytic virus} 
The virus is injected into the tumour, then diffuses in space with coefficient $D_v$ and decays with rate $q_v$. The lysis of an infected cell releases $\alpha$ viral particles. The dynamics are described by the equation
\begin{equation}
	\pt v(t,\vec{x})= \underbrace{D_v \Delta v(t,\vec{x})}_{\text{diffusion}} +\underbrace{\alpha q_I I(t,\vec{x})}_{\text{release}} -\underbrace{q_v v(t,\vec{x})}_{\text{natural decay}} + \underbrace{ v_{\text{inj}}(\vec{x}) \delta_{T_{\text{inj}}}(t)}_{\text{viral injection}}.
	\label{eq_v}
\end{equation}
The injection is modelled through a Dirac delta, which corresponds to a jump in the viral concentration at time $T_{\text{inj}}$; the spatial profile of the injection is given by the function $v_{\text{inj}}$.

\paragraph{Oxygen}
The oxygen is delivered by tissue vascularisation with a space-dependent intensity $Q(\vec{x})$, diffuses in space with diffusion coefficient $D_O$ and is consumed both by healthy tissue and cancer cells. The resulting equation is
\begin{equation}
	\pt O(t,\vec{x})= \underbrace{D_O \Delta O(t,\vec{x})}_{\text{diffusion}} - \underbrace{q_O O(t,\vec{x})}_{\text{natural decay}} -\underbrace{\lambda \rho(t,\vec{x}) O(t,\vec{x})}_{\text{cancer cell consumption}} + \underbrace{Q(\vec{x})}_{\text{source}}.
	\label{eq_O}
\end{equation}
We assume that healthy cells initially fill the tissue up to carrying capacity and their oxygen consumption is indirectly modelled through the decay at rate $q_O$. As the tumour grows, healthy cells are replaced by cancer cells, which consume more oxygen: the additional consumption is captured by the term $\lambda \rho(t,\vec{x}) O(t,\vec{x})$. Observe that both infected and uninfected cells are responsible for oxygen consumption in the same way and the epigenetic trait of uninfected cells does not affect the consumption (as already done in Ref. \cite{chiari23radio}).

\paragraph{Summary of the model}
Collecting all the equations together, the evolution of the system reads:
\begin{equation}
	\label{eq:complete}
	\begin{cases}
		\pt u(t,\vec{x},y)= D_y \pyy u(t,\vec{x},y) + D_{\vec{x}} \dive_{\vec{x}} (u(t,\vec{x},y) \nabla\rho(t,\vec{x}))\\
		\phantom{\pt u(t,\vec{x},y)=}+ P(y,\rho(t,\vec{x}))\; u(t,\vec{x},y)-S(y,O(t,\vec{x}))\,u(t,\vec{x},y)\\
		 \phantom{\pt u(t,\vec{x},y)=}-\beta(y)\; u(t,\vec{x},y)\; v(t,\vec{x}) \\
		\pt I(t,\vec{x})= D_{\vec{x}} \dive (I(t,\vec{x})\nabla \rho(t,\vec{x})) + v(t,\vec{x}) \int_Y \beta(y) u(t,\vec{x},y) \di y -q_I I(t,\vec{x}) \\
		\pt v(t,\vec{x})= D_v \Delta v(t,\vec{x}) +\alpha q_I I(t,\vec{x}) -q_v v(t,\vec{x})  + v_{\text{inj}}(\vec{x}) \delta_{T_{\text{inj}}(t)}\\
		\pt O(t,\vec{x})= D_O \Delta O(t,\vec{x}) -q_O O(t,\vec{x}) -\lambda \rho(t,\vec{x}) O(t,\vec{x}) + Q(\vec{x})\\
		\rho(t,\vec{x})\coloneqq \int_Y u(t,\vec{x},y)\di y +I(t,\vec{x})
	\end{cases}
\end{equation}
with the previously defined $\rho(t,\vec{x})$ in Eq. \eqref{eq_rho}, $P(y, \rho(t,\vec{x}))$ in Eq. \eqref{eq_P}, $S(y,O(t,\vec{x}))$ in Eq. \eqref{eq_S} based on $\varphi(O(t,\vec{x}))$ in Eq. \ref{eq_phi}, and $p(y)$ and $\beta(y)$ set as in Eq. \eqref{eq_pbeta}. We keep the oxygen source $Q(x)$ in general form and change it according to the biological setting we aim to reproduce.

We define the Cauchy problem by imposing the initial conditions: 
\begin{equation*}
	\begin{cases}
		u(0,\vec{x},y)= u_0(\vec{x},y)\\
		I(0,\vec{x})= 0\\
		v(0,\vec{x})= 0\\
		O(0,\vec{x})= O_0(\vec{x})
	\end{cases}
\end{equation*}
where $u_0(\vec{x},y)$ and $O_0(\vec{x})$ will be defined in the context of the various scenarios in Section \ref{sec:results}; we always assume that the tumour initially grows without viral infection and the therapy is administered after some time. Moreover, we impose no flux boundary conditions on $\partial Y$, i.e. $\partial_{y}u(t,\vec{x},0)=\partial_{y}u(t,\vec{x},1)=0$,
corresponding to the fact that the epigenetic trait cannot assume values below $0$ or above $1$. Finally, we also impose no flux boundary condition for all $u(t,\vec{x},y)$, $I(t,\vec{x})$, $v(t,\vec{x})$ and $O(t,\vec{x})$ at $\partial\Omega$, meaning that these quantities cannot leave the spatial domain.

\subsection{Theoretical insights}
\label{sec:theory}

The stationary equilibrium values of the system described in Eq. \eqref{eq:complete} in the spatially homogeneous case can be computed through formal asymptotic analysis.
Building upon the methods employed in Refs. \cite{lorenzipainter22,lorenzi21invasion,villa21modeling}, we introduce a small parameter $\epsilon$ and assume that $D_y=\epsilon^2$. Furthermore, we use the time scaling $t\mapsto \frac{t}{\epsilon}$, which allows us to study the long-time behaviour of the system. 
We then make for uninfected cells the real phase WKB ansatz \cite{barles90,evans89,fleming86}, as it is common in the Hamilton--Jacobi approach presented, for example, in \cite{diekmann05,lorz11,perthame2006transport,perthame08}
\[
u_\epsilon(t,\vec{x},y)= e^{\frac{n_\epsilon(t,\vec{x},y)}{\epsilon}}.
\]
Furthermore, we assume that, for every $\epsilon>0$, the initial condition $n_\epsilon(0,\vec{x},y)$ is a uniformly concave function of $y$; for example, we could think that, at $t=0$, tumour cells that occupy the same position are mainly in the same phenotypic state and so the local cell phenotypic distribution  $y\mapsto u_\epsilon(0,\vec{x},y)$ is a sharp Gaussian-like function (meaning that $n_\epsilon(0,\vec{x},y)=-(y-\varphi_\epsilon(\vec{x}))^2$, for some appropriate function $\varphi_\epsilon$). Since $y\mapsto R(y,\rho,I)$ is a concave function of $y$, we also expect $n_\epsilon$ to be a strictly concave function of $y$ \cite{mirrahimi2015asymptotic, perthame08}. We can then define
\begin{equation}
\label{eq:def_ybar_spatial}
\bar{y}_\epsilon(t,\vec{x})\coloneqq \arg\max_{y\in Y} n_\epsilon(t,\vec{x},y).
\end{equation}


\david{For $\vec{x}\in\supp(\rho)$ and $\epsilon\to 0$ we have
\begin{equation}
\label{eq:selectedt}
u(t,\vec{x},y)\overset{*}{\rightharpoonup} U(t,\vec{x}) \delta_{\bar{y}(t,\vec{x})}
\end{equation}
in the sense of weak-$*$ convergence of measures; $\delta_{\bar{y}(t,{x})}$ is the Dirac delta distribution centred at $\bar{y}(t,\vec{x})$, which is the fittest epigenetic trait at time $t$ and position $\vec{x}$. This is a consequence of the boundedness of $R$, as explained in \ref{app:theory}. }

The computations performed in \ref{app:theory} show that, if the system converges as $\epsilon\to 0$ to a spatially homogeneous steady state, which we denote by $(U^\infty,I^\infty,v^\infty,O^\infty,\bar{y}^\infty)$, then \david{both $R$ and $\py R$ vanish at the equilibrium. The equilibrium values also solve the stationary version of Eq. \eqref{eq:complete} and in the case of a persistent infection these conditions lead to the system}
\begin{equation}
	\label{eq:equilibria_complete}
	\begin{cases}
		U^\infty=\dfrac{q_v}{\alpha [\beta_M+(\beta_m-\beta_M)\bar{y}^\infty]} \\[8pt]
		v^\infty=\dfrac{\alpha q_I}{q_v} \, I^\infty \\[8pt]
		O^\infty=\dfrac{Q}{q_O+\lambda(U^\infty+I^\infty)}\\[8pt]
		R(\bar{y}^\infty,U^\infty+I^\infty,O^\infty,v^\infty) =[p_M+(p_m-p_M)\bar{y}^\infty] \Bigl( 1-\dfrac{U^\infty+I^\infty}{K}\Bigr) \\
        \phantom{R(\bar{y}^\infty,U^\infty+I^\infty,O^\infty,v^\infty) =}- \eta (\bar{y}^\infty-\varphi(O^\infty))^2 \\
		\phantom{R(\bar{y}^\infty,U^\infty+I^\infty,O^\infty,v^\infty) =}-[\beta_M+(\beta_m-\beta_M)\bar{y}^\infty] v^\infty=0\\
		\py R(\bar{y}^\infty,U^\infty+I^\infty,O^\infty,v^\infty)=(p_m-p_M)\Bigl( 1-\dfrac{U^\infty+I^\infty}{K}\Bigr) \\
		\phantom{\py R(\bar{y}^\infty,U^\infty+I^\infty,O^\infty,v^\infty) =}- 2\eta (\bar{y}^\infty-\varphi(O^\infty)) -(\beta_m-\beta_M) v^\infty=0 \\[8pt]
	\end{cases}
\end{equation}
In \ref{app:theory}, we also discuss the case of infection-free equilibria.

The above system is too complicated to be studied analytically; hence, we mainly consider numerical solutions. Although six solutions exist, only one is biologically meaningful in the parameter range that we consider. 
A more useful expression can be obtained by solving the equation $\py R=0$ in $\bar{y}^\infty$:
\begin{equation}
	\label{eq:ybar}
	\bar{y}^\infty=\varphi(O^\infty) + \frac{1}{2\eta}\Bigl[-(p_M-p_m)\Bigl(1-\frac{\rho^\infty}{K}\Bigr)+(\beta_M-\beta_m)v^\infty\Bigr].
\end{equation}
This formula allows us to understand how the fittest trait $\bar{y}^\infty$ is influenced by the environmental conditions, the competition among cell types and the infection. Indeed,
$\varphi(O^\infty)$ is the epigenetic trait that is favoured from the oxygen-induced selective pressure. The actual fittest trait $\bar{y}^\infty$ tends to decrease with respect to $\varphi(O^\infty)$ when the total cell population $\rho$ is low due to the different proliferation rates of cell lines in a situation of low competition; as the total cell density approaches carrying capacity, the growth slows down for all epigenetic traits and the difference in the proliferation rates becomes less relevant. The fittest trait $\bar{y}^\infty$ is also affected by the viral infection: it grows in the presence of viral particles that targets proliferative cells and it reduces if the infection targets hypoxic cells. We remark that this formula may yield a value of $\bar{y}^\infty$ outside the interval $[0,1]$, which has no meaning in our formulation of the model: indeed, \david{the condition $\py R=0$ is only verified if $\py n(t,\bar{y}(t))=0$, which is true for $\bar{y}\in (0,1)$ (see also \ref{app:theory}).} The two boundary cases can be easily treated separately: when the previous equations yield a value $\bar{y}^\infty<0$, we should expect the fittest trait to be $0$; when the previous equations yield a value $\bar{y}^\infty>1$, we should expect the fittest trait to be $1$.

\begin{figure}
	\centering
	\includegraphics[width=0.75\linewidth]{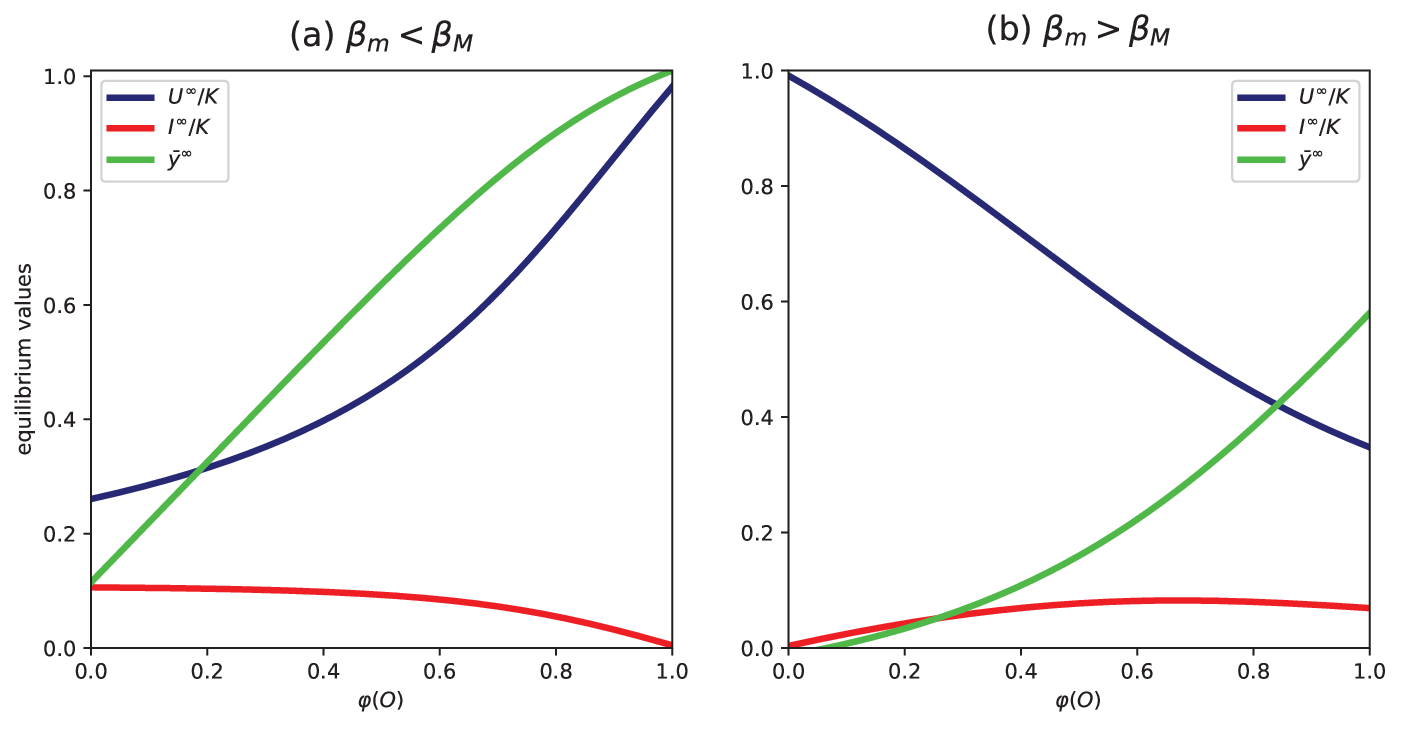}
	\caption{Numerical solution of Eq. \eqref{eq:equilibria_hyp} showing the equilibria in different oxygen conditions. The parameters in panel (a) take the values listed in Table \ref{tab:parameters}. In panel (b), the values of $\beta_M$ and $\beta_m$ are switched to reproduce the situation of oncolytic viruses that specifically target hypoxic cells. In both cases, $\varphi(O^\infty)$ ranges between $0$ and $1$.}
	\label{fig:equilibria}
\end{figure}

A simpler situation is obtained by assuming that tumour dynamics do not significantly affect oxygen density so that $O^\infty$ is given a priori; in this case, the system becomes
\begin{equation}
	\label{eq:equilibria_hyp}
	\begin{cases}
		U^\infty=\dfrac{q_v}{\alpha [\beta_M+(\beta_m-\beta_M)\bar{y}^\infty]} \\[8pt]
		v^\infty=\dfrac{\alpha q_I}{q_v} \, I^\infty \\[8pt]
		R(\bar{y}^\infty,U^\infty+I^\infty,O^\infty,v^\infty) =[p_M+(p_m-p_M)\bar{y}^\infty] \Bigl( 1-\dfrac{U^\infty+I^\infty}{K}\Bigr) \\
       \phantom{R(\bar{y}^\infty,U^\infty+I^\infty,O^\infty,v^\infty) =} - \eta (\bar{y}^\infty-\varphi(O^\infty))^2 \\
		\phantom{R(\bar{y}^\infty,U^\infty+I^\infty,O^\infty,v^\infty) =}-[\beta_M+(\beta_m-\beta_M)\bar{y}^\infty] v^\infty=0\\
		\py R(\bar{y}^\infty,U^\infty+I^\infty,O^\infty,v^\infty)=(p_m-p_M)\Bigl( 1-\dfrac{U^\infty+I^\infty}{K}\Bigr)  \\
		\phantom{\py R(\bar{y}^\infty,U^\infty+I^\infty,O^\infty,v^\infty)=}- 2\eta (\bar{y}^\infty-\varphi(O^\infty))-(\beta_m-\beta_M) v^\infty=0 \\[8pt]
	\end{cases}
\end{equation}
It is then possible to obtain a third-degree equation for $\bar{y}^\infty$, which in principle can be solved; however, the explicit solutions are still too complicated to give any useful information. Fig. \ref{fig:equilibria} shows the numerical solution of Eq.~\eqref{eq:equilibria_hyp} that is biologically meaningful. In the reference situation ($\beta_m<\beta_M$), the equilibrium values of $U^\infty$ and $\bar{y}^\infty$ increase as the oxygen values decrease; when the values of $\beta_M$ and $\beta_m$ are switched, we observe the inverse behaviour. The effect of oxygen variations on $I^\infty$ is more complex, as its value is almost constant for a wide range of oxygen values and then significantly decreases only when the oxygen concentration is very low (or very high).

\begin{table}[t!]
	\centering{
		\scriptsize
		\begin{tabular}{llp{0.3\linewidth}ll}
			&\multicolumn{1}{c}{\textbf{Parameter}} & \multicolumn{1}{c}{\textbf{Description}} &  \multicolumn{1}{c}{\textbf{Value [Units]}} & \multicolumn{1}{c}{\textbf{Reference}}\\
			\toprule
			& $p_M$         &maximal duplication rate      & $2.88 \times 10^{-2}$ [h$^{-1}$] &\cite{ke00} \\
			& $p_m$         &minimal duplication rate      & $1.44 \times 10^{-2}$ [h$^{-1}$] &\cite{martinez12} \\
			& $K$         &tissue carrying capacity    & $10^6$ [cells/mm$^3$]	&\cite{lodish08} \\
			& $D_{\vec{x}}$         &cell spatial diffusion coefficient    & $1.30\times 10^{-9}$ [(mm$\times$cells$\times$h)]	&estimate based on \cite{kim06} \\
			& $\eta$         &selection rate by oxygen     & $2.08 \times 10^{-2}$ [h$^{-1}$] &model estimate \\
			& $D_y$         &cell epigenetic diffusion coefficient    & $5.00\times 10^{-6}$ [h$^{-1}$]]	&\cite{celora21} \\
			\midrule
			& $\beta_M$         &maximal infection rate   & $7.00\times 10^{-10}$ [mm$^3$/(viral particles$\times$h)]	&\cite{friedman06} \\
			& $\beta_m$         &minimal infection rate   & $1.75\times 10^{-10}$ [mm$^3$/(viral particles$\times$h)]	&model estimate \\
			& $q_I$         &death rate of infected cells     & $4.17\times 10^{-2}$ [h$^{-1}$]	&\cite{ganly00} \\
			& $q_v$ 			&virus clearance rate		&$1.67\times 10^{-1}$ [h$^{-1}$] &\cite{mok09} \\
			& $\alpha$         &viral burst size     & $1000$ [viral particles/cells]	&model estimate \\
			& $D_v$         &virus diffusion coefficient    & $3.6\times 10^{-2}$ [mm$^2$/h]	&\cite{kim06} \\
			\midrule
			& $O_{\max}$         &maximal oxygen concentration     & $2.16\times 10^{-3}$ [mm$_{O_2}^3$/mm$^3{_\text{plasma}}$]	&\cite{mckeown14,pittman11}\\
			& $O_M$         &oxygen normoxic threshold     & $1.71\times 10^{-3}$ [mm$_{O_2}^3$/mm$^3{_\text{plasma}}$]	&\cite{mckeown14,pittman11}\\
			& $O_m$         &oxygen hypoxic threshold     & $2.28\times 10^{-4}$ [mm$_{O_2}^3$/mm$^3{_\text{plasma}}$]	&\cite{mckeown14,pittman11}\\
			& $q_O$         &oxygen physiological decay rate     & $5.60\times 10^{2}$ [h$^{-1}$]	&estimate based on \cite{wagner11} \\
			& $\lambda$         &oxygen consumption rate     & $6.55 \times 10^{-4}$ [mm$^3$/(cell$\times$h)]	&estimate based on \cite{grimes14}\\
			& $D_O$         &oxygen diffusion coefficient   & $3.60$ [mm$^2$/h]	&\cite{mueller-klieser84} \\
			\bottomrule
		\end{tabular}
		\caption{Parameter set.}\label{tab:parameters}}
\end{table}

So far, we have focused our discussion on the spatially homogeneous situation. To our knowledge, spatial dynamics in this context have not been studied analytically, not even in simpler settings (such as the one of Ref.~\cite{lorenzi21}). Nevertheless, given the forthcoming numerical simulations, it is helpful to recall some elementary facts, \david{with the goal of elucidating several complex dynamics that we observe}. If we neglect the epigenetic structure, the one-dimensional dynamics of uninfected cells in the absence of viral infection follow the equation
\[
\pt u(t,{x})=D_{\vec{x}} \px (u(t,{x}) \px u(t,{x}))+ p\Bigl(1-\frac{u(t,{x})}{K}\Bigr) u(t,{x}).
\]
It is well-known that there exist travelling waves solutions of this equation with speed at least $\sqrt{D_{\vec{x}}Kp/2}$ and an initial condition with compact support evolves into a wave that travels with the minimal speed \cite{aronson80,newman80}. The addition of infection due to cell-to-cell contact originates travelling waves of the two populations (uninfected and infected), with the uninfected proliferative cells trying to escape from the infected cells \cite{morselli23}; the addition of a viral population does not significantly change this picture. It is important to remark that an efficient infection results in a wavefront much lower than the carrying capacity, whose invasion speed is lower than $\sqrt{D_{\vec{x}}Kp/2}$.

Such results do not directly apply to Eq. \eqref{eq:complete} because of the presence of the epigenetic structure. A rough estimate can be obtained by integrating Eq. \eqref{eq_U} in $y$ and assuming $u(t,{x},y)= U(t,{x}) \delta_{\bar{y}(t,{x})}$; we also neglect viral infection at the invasion front. As a result we obtain the equation
\begin{align*}
\pt U(t,{x})=&D_{\vec{x}} \px (U(t,{x}) \px U(t,{x}))+ \Bigl[p(\bar{y}(t,{x}))\Bigl(1-\frac{U(t,{x})}{K}\Bigr) \\
&-\eta(\bar{y}(t,x)-\varphi(O))^2 \Bigr] U(t,{x}).
\end{align*}
If we had $\bar{y}(t,{x})\approx \varphi(O)$ with $O$ constant in time and space, then it is clear that the minimal wave speed would be $\sqrt{D_{\vec{x}}Kp(\varphi(O))/2}$. The numerical simulations of the next section show that this is not the case, not even when the oxygen concentration is stationary and homogeneous: indeed, a rigorous study of the travelling waves should keep into account the spatial variations of $\bar{y}$ (see Ref. \cite{lorenzi24preprint} and the references therein), which are extremely complex the case of our system and fall beyond of the scope of the present work; further complexities arise from the fact that the invasion front is lower than $K$ because of the infection and this also affects the wave speed. Nevertheless, the previous considerations provide us a broad intuition on the behaviour of the system, which is in line with the forthcoming numerical simulations.

\section{Numerical results}
\label{sec:results}

In this Section, we compare numerical simulations with the theoretical results described in the previous Section. 
To make the comparison between the epigenetic composition and spatial characterisation more straightforward, we introduce the \textit{average epigenetic trait}, defined as
\[
\mu(t,\vec{x}) \coloneqq \frac{\int_Y u(t,\vec{x},y)\,y\di y}{U(t,\vec{x})} \qquad \text{for all } (t,\vec{x}) \text{ such that } U(t,\vec{x})\neq 0.
\]
Observe that this condition is not satisfied in the whole domain due to the compact support of $u$.
The relevance of this quantity comes from the fact that, in general, we do not expect the full selection of a single trait in our numerical simulations due to the epigenetic diffusion. While it is still possible to define the \textit{most frequent epigenetic trait} as in Eq. \eqref{eq:def_ybar_spatial} (which, in light of the asymptotic analysis, we expect to be well-defined whenever $U(t,\vec{x})\neq 0$), one should also take into account that its numerical approximation is constrained by definition on the discrete mesh of $Y$ and may provide a good estimate of the theoretical value only if the mesh is very small. This issue can be easily solved by considering instead the average epigenetic trait, as defined above. For $u(t,\vec{x},y)=U(t,\vec{x}) \delta_{\bar{y}(t,\vec{x})}$, clearly $\mu(t,\vec{x})$ coincides with $\bar{y}(t,\vec{x})$. 
In general, it is still reasonable to expect the epigenetic distribution to be approximately a Gaussian function concentrated around a maximum value and so in theory $\bar{y}(t,\vec{x})\approx\mu(t,\vec{x})$; nevertheless, the numerical approximation of the quantity $\mu$ is more robust than the one of $\bar{y}$. Hence we focus our analysis on $\mu$, keeping in mind that the asymptotic results related to $\bar{y}$ (including Eq. \eqref{eq:ybar}) apply equally well for $\mu$.

The parameter values are listed in Table \ref{tab:parameters}; in \ref{app:num}, we explain how the values have been chosen, as well as the numerical method employed. 
In all the simulations, we start with an uninfected tumour of the form
\begin{equation}
	\label{eq:u_init}
	u_0(\vec{x},y) = 
	\begin{cases}
		A_u\; e^{-\frac{|\vec{x}-\vec{x}_0|^2}{\theta_{\vec{x}}}-\frac{(y-y_0)^2}{\theta_y}} \quad &\text{if } A_u\; e^{-\frac{|\vec{x}-\vec{x}_0|^2}{\theta_{\vec{x}}}-\frac{(y-y_0)^2}{\theta_y}}>1 \\[8pt]
		0 \quad &\text{otherwise}
	\end{cases}
\end{equation}
The truncation is performed in order to have an initial condition with compact support; the form of the equations is such that the solution will still be compactly supported at all times \cite{aronson80,newman80}. In all the simulations we set $\vec{x}_0=(0,0)$, $y_0=\varphi(O(0,\vec{x}_0))$, $\theta_{\vec{x}}=0.5$, $\theta_{y}=0.5$. The parameter $A_u$ is set to
\[
A_u=\frac{7.19\times 10^4\;\text{cell/mm}^3}{\displaystyle\int_{Y}e^{-\frac{(y-y_0)^2}{\theta_y}} \di y}.
\]
This choice allows a maximal initial total cell density equal to $\frac{K}{10}$, irrespectively of the value of $y_0$.
We then assume that viral injection is performed after some time so that the tumour can adapt to the environment. In most of the cases, we perform a central viral injection as soon as the tumour reaches a given size: in mathematical terms, we set
\begin{equation}
	\label{eq:T_inj}
	T_{\text{inj}}\coloneqq \inf \Set{t\in [0,+\infty) | d(t) \geq d_{\text{inj}}}
\end{equation}
where $d_{\text{inj}}$ is the tumour size at which we choose to inject the virus and
\[
d(t)\coloneqq \diam \Set{\vec{x}\in\Omega | \rho(\vec{x},t)\geq \frac{K}{10}}
\]
We recall that the diameter of a general set $E$ is defined as 
\[
\diam E \coloneqq \sup \Set{\abs{\vec{x}_1-\vec{x}_2} | \vec{x}_1,\vec{x}_2 \in E}.
\]
In the particular situation of a circle, this definition coincides with the standard diameter; in the general case, the diameter is the longest length found inside the set. This choice is based on the assumption that small tumours cannot be clinically detected, hence the therapy may only start when cancer cells reach a density of at least one tenth of the carrying capacity in a large region. We set $d_{\text{inj}}=5.2\;$mm, as in Ref. \cite{kim06} and the central viral injection takes the form
\begin{equation}
	\label{eq:v_init}
	v_{\text{inj}}(\vec{x}) = A_v\; e^{-\frac{|\vec{x}-\vec{x}_0|^2}{\theta_v}}
\end{equation}
with $A_v=7\times 10^9$, $\theta_v=0.5$. This allows for a total number of viral particles in agreement with the experiments performed in Ref. \cite{kim06}. 

\begin{table}[ht]
\footnotesize
\centering
\giulia{
\makebox[0pt][c]{
\begin{tabularx}{0.9\textwidth}{l|Z|Z|l|Y|}
\cline{2-5}
& \multicolumn{2}{c|}{\textbf{Oxygen}} & \textbf{Virotherapy} & \textbf{Aim} \\ 
\cline{2-3} 
& space & time &  &   \\
\cline{1-5}
\multicolumn{1}{|l|}{\expzero} & homog.& const. & none &  oxygenation-dependent effects on tumour evolution\\ \cline{1-5}
\multicolumn{1}{|l|}{\exponea} & homog. & const.  & std. & oxygenation-dependent effect on therapy \\ \cline{1-5}
\multicolumn{1}{|l|}{\exponeb} & heterog. & const. & std. & spatial heterogeneity effects on treated tumours\\ \cline{1-5}
\multicolumn{1}{|l|}{\exptwoa} & heterog. & dyn. & std. & co-dependence of time-dependent dynamics  \\ \cline{1-5}
\multicolumn{1}{|l|}{\exptwob} & heterog. & dyn. & hyp. spec. & alternative therapeutical approach \\ \cline{1-5}
\end{tabularx}
}
\caption{Schematic description of experiments in Sections \ref{subsec:evolv}, \ref{subsec:stat} and \ref{subsec:dyn}.}
\label{tab:experiments}
}
\end{table}

\giulia{We now provide a roadmap outlining the logical flow underlying the proposed sequence of experiments (for a schematic summary, see Table \ref{tab:experiments}).}
For the sake of simplicity, we first consider 
\giulia{the natural evolution of the tumour in the absence of therapy and with constant oxygenation in space and time, corresponding to a situation in which the tumour does not influence the oxygen distribution and the environment is homogeneous (Section \ref{subsec:evolv}). While this is clearly an oversimplification, it allows us to focus our attention on the tumour's evolutionary dynamics and the direct effect of oxygen concentration. 
We then add the therapy (Section \ref{subsec:stat}), considering standard oncolytic viruses, to study the effect of oxygen concentration on this treatment strategy. Moreover, we add a degree of complexity allowing for spatially-inhomogeneous oxygen sources.
 Finally we analyse the full model (Section~\ref{subsec:dyn}), which includes oxygen dynamics, taking into account different configurations of oxygen sources. We adopt this complete scenario to also mention the case of a virus that specifically infects hypoxic cells, looking towards the combination of oncolytic virotherapy with other treatments.}

\subsection{\giulia{Therapy-free tumour evolution with minimal oxygen characterisation (\expzero)}}
\label{subsec:evolv}
\begin{figure}
	\centering
	\includegraphics[width=0.75\linewidth]{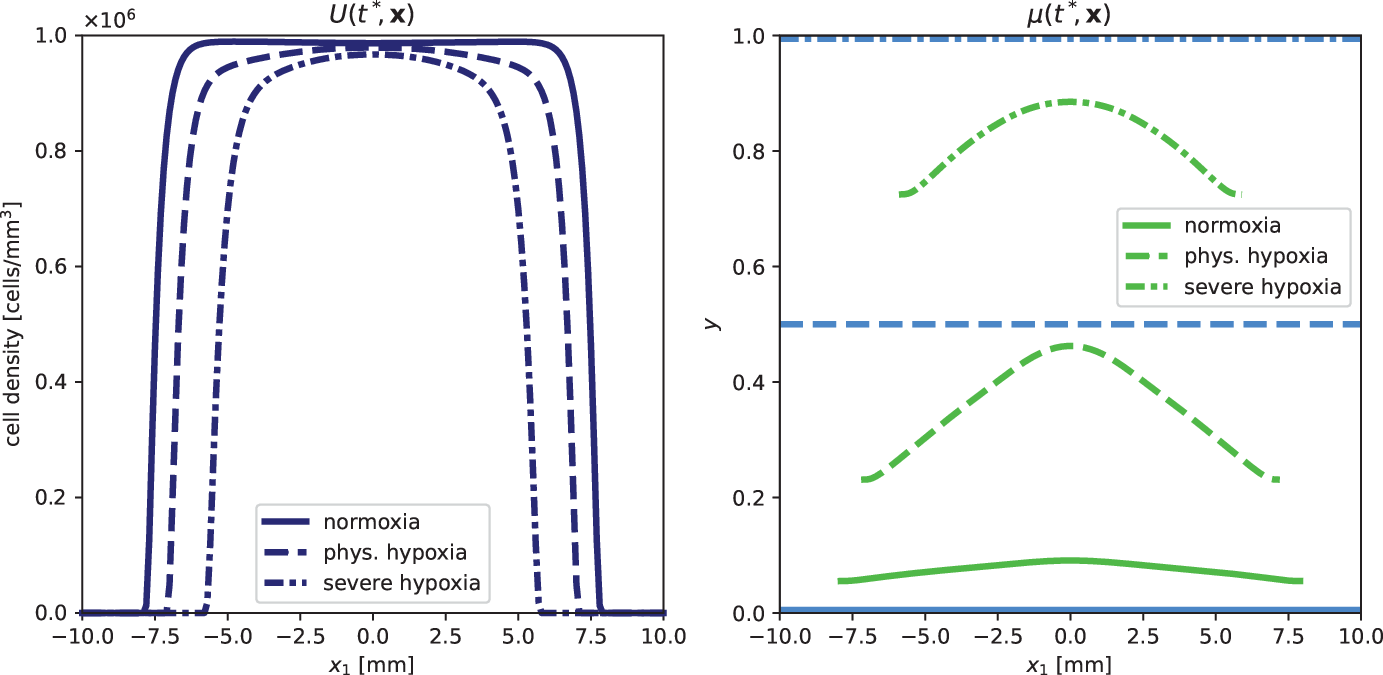}
	\caption{Results of the numerical simulation without viral infection for stationary oxygenation, for three spatially homogeneous oxygen condition: $O=O_M$ (solid lines), $O=\frac{O_M+O_m}{2}$ (dashed lines), and $O=O_m$ (dot-dashed lines). We plot the solutions at time $t^*=1500\;$h. Only the horizontal section is shown to facilitate the comparison. The blue lines in the left plot represent the profile of uninfected cancer cells $U(t^*,\vec{x})$. In the right plot, the green lines show the average epigenetic trait $\mu(t^*,\vec{x})$ and the light-blue lines show the fittest trait selected by the environment $\varphi(O(t^*,\vec{x}))$.}
	\label{fig:hom_stat_overlap}
\end{figure}

As a starting point, it is helpful to observe how a tumour evolves \giulia{in the context of a basic oxygenation layout, with no heterogeneity in space and time, and } without treatment. \giulia{This preliminary case  helps us to understand which dynamics can be attributed solely to natural evolution, thereby providing a baseline to better interpret, in subsequent cases, which observations result from the effects of therapy and the spatial and temporal heterogeneity of oxygenation. We focus on three oxygen values, namely $O_M$ (normoxia), $O_m$ (severe hypoxia) and their average $\frac{O_m+O_M}{2}$ (physiological hypoxia), whose corresponding selected traits are respectively $1$, $0$ and $0.5$; other values produce intermediate situations.

Results are represented in} Fig. \ref{fig:hom_stat_overlap} for $t=1500\;$h (for the sake of clarity, the figure represents the central section of the domain, i.e. the set $[-L,L]\times\{0\}$). The left panel shows the total uninfected cell density $U(t,\vec{x})$, which in absence of infection coincides with the total cancer density $\rho(t,\vec{x})$; the right panel shows the average epigenetic trait $\mu(t,\vec{x})$. Overall, we observe the behaviour predicted by the theoretical asymptotic analysis in all the cases. The three initial conditions are given by Eq. \eqref{eq:u_init} with $y_0=\varphi(O)$. However, the density $\rho$ at the beginning is much lower than $K$; hence, unless we are in a normoxic situation, the fittest epigenetic trait is lower than $\varphi(O)$, as predicted by Eq. \eqref{eq:ybar}; on the other hand, in the normoxic situation $\varphi(O)=0$ is already the lowest attainable value. As time passes, the cell density grows close to carrying capacity and the cancer starts to invade the surrounding area. In the hypoxic scenarios, the fittest epigenetic trait grows with $\rho$ until reaching the value $\varphi(O)$; however, that trait is never completely selected due to epigenetic diffusion. An important consequence of the presence of different epigenetic characteristics is the fact that $\rho$ is always slightly below $K$, as the oxygen-driven selective pressure never completely stops: this effect is especially evident in the hypoxic situation, in which the slow proliferation contrasts the selective pressure less effectively. \david{In all cases,} the average epigenetic traits are lower at the invasion front, due to the lower total densities (see Eq. \eqref{eq:ybar}), and increase as we get close to the tumour centre. Overall, high oxygen levels are associated with more proliferative tumours, which reach carrying capacity earlier and invade the surrounding tissues faster \david{(in line with the theoretical value of the wave speed)}.

\subsection{\giulia{The stationary oxygen approximation: standard oncolytic viruses}}

\giulia{Following a stepwise approach, we gradually introduce the elements of complexity in the model. In this section, we focus on the influence of oncolytic virotherapy, limiting our analysis to the case of a standard virus and assuming constant oxygenation over time. Dynamics converge to the theoretical equilibrium faster than in the case of the full model; as a consequence, this situation allows us to test the reliability of the asymptotic analysis.

In the two subsequent subsections, we first consider a spatially homogeneous oxygenation (Section \ref{subsubsec:case1a}) and then introduce spatial heterogeneity (Section \ref{subsubsec:case1b}). From a biological perspective, both experiments investigate how the tumour dynamics respond to the action of oncolytic viruses under different environmental conditions. In the second case, in particular, we examine the effects of environmental heterogeneity on the geometric heterogeneity of the tumour expansion rate and on the spatially-specific composition of the mass in terms of quantitative measures, epigenetic characteristics, and infection.}
\label{subsec:stat}

\subsubsection{\giulia{Homogeneous oxygen distribution (\exponea)}}
\label{subsubsec:case1a}
\begin{figure}
	\centering
	\includegraphics[width=0.75\linewidth]{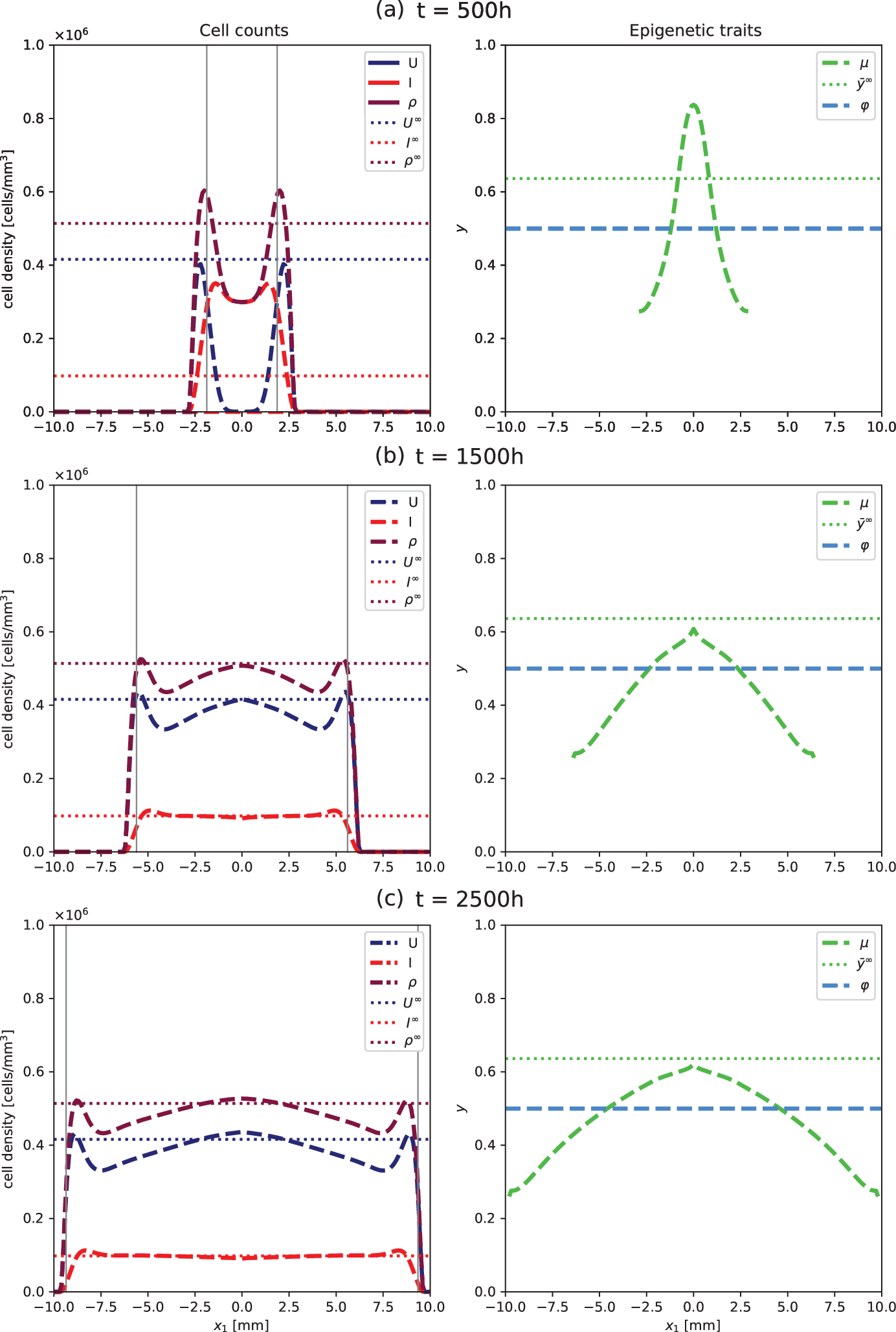}
	\caption{Results of the numerical simulations for stationary oxygen under physiological hypoxia, at times $t=500\;$h (panel (a)), $t=1500\;$h (panel (b)), and $t=2500\;$h (panel (c)). First column shows $U(t^*,\vec{x})$ in blue, $I(t^*,\vec{x})$ in red, and $\rho(t^*,\vec{x})$ in purple. The dotted lines show the theoretical approximation of asymptotic equilibria (using correspondent colours), obtained by solving Eq. \eqref{eq:equilibria_hyp}.
    The second column provides the average epigenetic trait $\mu(t^*,\vec{x})$ in green and $\varphi(O(t^*,\vec{x}))$ in light blue.  Vertical grey lines indicate the position of a wave-front moving with speed $\sqrt{D_{\vec{x}}Kp(\varphi(O))/2}$ (with $\varphi(O)$ equal respectively to $0$, $0.5$ and $1$).}
	\label{fig:time_evolution}
\end{figure}

\begin{figure}
	\centering
	\includegraphics[width=0.75\linewidth]{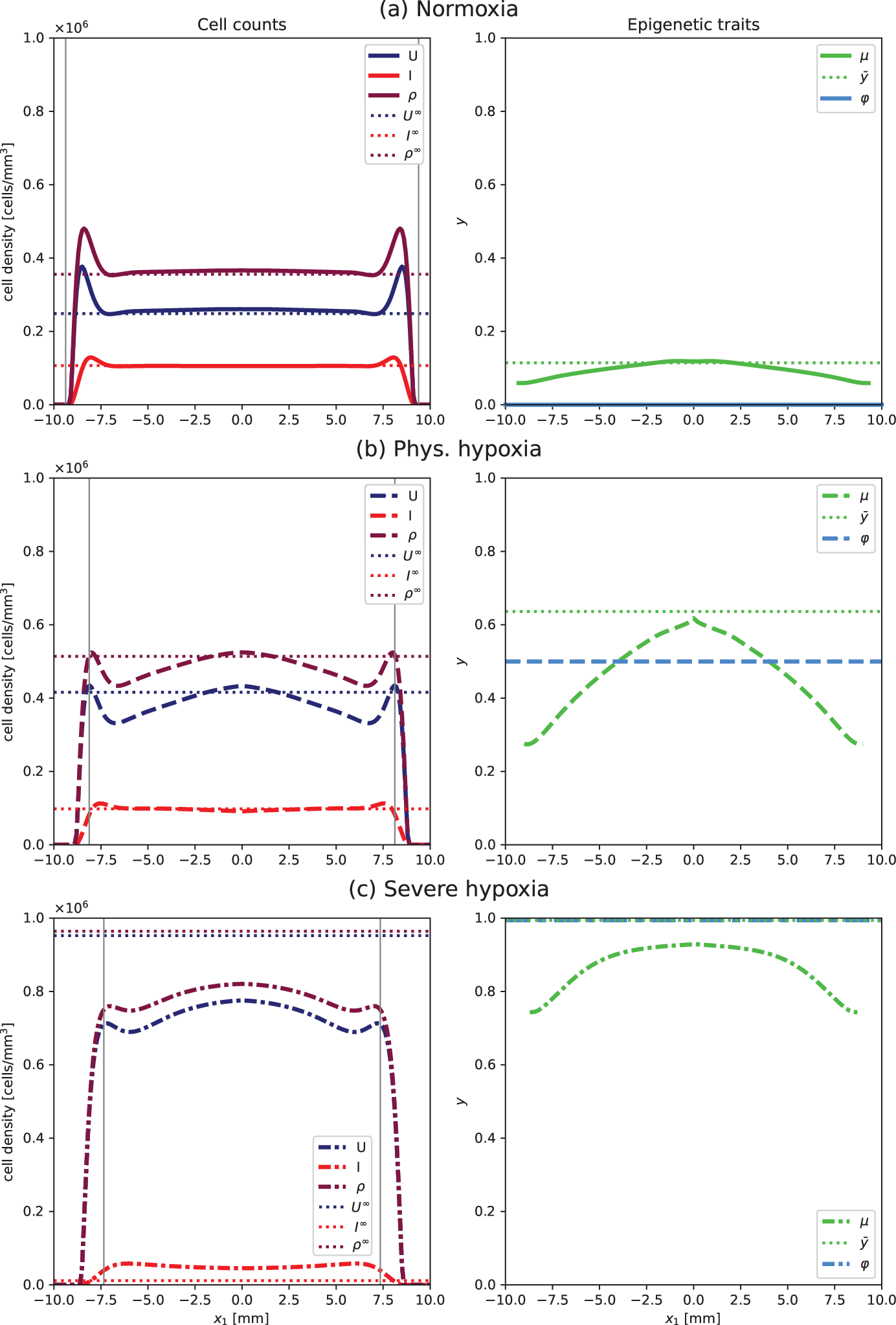}
	\caption{Results of the numerical simulations for stationary oxygen at time $t^*$, corresponding to approximately $1800\;$h after viral injection, for three spatially homogeneous oxygen conditions: $O=O_M$ (solid lines, panel (a)), $O=\frac{O_M+O_m}{2}$ (dashed lines, panel (b)), and $O=O_m$ (dot-dashed lines, panel (c)). All the graphical elements have the same meaning as in Fig. \ref{fig:time_evolution}. Only the horizontal section is shown to facilitate the comparison. Note that $\bar{y}(t^*,\vec{x})=1$ in panel (c) (i.e., the maximum value is on the boundary of $Y$).}
	\label{fig:hom_stat_comparison}
\end{figure}

Even though from the previous discussion hypoxic tumours may seem less aggressive (at least in terms of growth velocity) the situation overturns in the presence of treatment, as the adaptation to hypoxia influences the infectivity, making the tumour less susceptible to therapies in the specific case of standard oncolytic viruses here considered. We again consider the oxygen values corresponding to normoxia, physiological hypoxia and severe hypoxia.

Let us begin with a general representation of the tumour evolution in presence of treatment. In Fig. \ref{fig:time_evolution} we consider the physiological hypoxia setting and we represent three time steps of tumour evolution with virotherapy (for the related complete video, see electronic supplementary material S2). 
In this setting, the tumour reaches a treatable size at around $t=471\;$h, thus panel (a) represents a time step in which the injection has just occurred. The total population is significantly smaller than the one reached at the moment of injection and centre of the tumour is characterised by a lower concentration of cells, almost all infected. At the represented instant, the infection is approaching the edges of the tumour, where uninfected cells are predominant and the average epigenetic trait is considerably lower. This is because the trait is still predominantly determined by environmental condition rather than by the selective pressure operated by the treatment. Panel (b) considers a time which follows the treatment and highlights how the central region of the tumour is quickly infected with viral particles able to reach the tumour front in a relatively short time due to their fast diffusion. As time progresses (see also panel (c)), cell densities at the centre of the tumour converge with damped oscillations to the equilibrium predicted by the theoretical analysis. The average of the epigenetic traits in the central area sensibly increases right after the viral injection, then oscillates towards the equilibrium. It is interesting to observe that epigenetic traits at the invasion front are lower. This is coherent with Eq. \eqref{eq:ybar} since, in this area,  $I$ is lower (in other words, tumour invasion is guided by the most proliferative cells) and $\rho$ is smaller too. It is notable to observe how wave fronts at the speed $\sqrt{D_{\vec{x}}Kp(0.5)/2}$ reproduce, qualitatively and quantitatively with good coherence, the advancement of the tumour mass.

We now proceed to compare different oxygenation scenarios, to understand in more detail how this factor influences the evolutionary dynamics of the tumour mass in the presence of therapy. Indeed, Fig. \ref{fig:hom_stat_comparison}, along with the video accompanying it (see electronic supplementary material S2), shows the effect of oncolytic virotherapy on the tumours in the three oxygenation settings introduced at the beginning of this section. The different growth rates imply that the tumours reach the threshold size $d_{\text{inj}}$ for treatment at different times; thus, the viral injection is performed respectively at around $t=426\;$h for normoxia, $t=471\;$h for physiological hypoxia and $t=609\;$h for severe hypoxia. To facilitate the comparison between the different scenarios, Fig. \ref{fig:hom_stat_comparison} shows the section of the simulation approximately $1800\;$h after the viral injection. \david{The virus quickly infects the whole tumour in all three cases}. In the severely hypoxic case, this initial successful infection might appear surprising. Still, it can be easily explained by the fact that at $T_{\text{inj}}$ the tumour has not reached the carrying capacity yet and the epigenetic characteristics are still not fully adapted to the environment (the lack of complete adaptation is also true in the other cases, but less evident).



\david{The normoxic and physiologically hypoxic situations can be described as partial successes of the therapy, in line with the equilibrium values predicted by the asymptotic analysis. On the other hand,} the severely hypoxic case (Fig. \ref{fig:hom_stat_comparison}c) is clearly a complete failure: the tumour density decreases only for a short time, after which it starts to regrow up to around $90\%$ of carrying capacity, with a very small fraction of infected cells (not shown here, see electronic supplementary material S2); we remark that such a low number of cells may model a situation of extinction due to stochastic events. After approximately $400\;$h, there is a relapse of the infection, which causes a small decrease in the total cell density followed by a subsequent regrowth towards the theoretical equilibrium. The convergence is extremely slow and it is clear from Fig. \ref{fig:hom_stat_comparison}c that $1800\;$h after the viral injection the dynamics are still far away from the equilibrium. This can be explained by the fact that the smaller growth rate slows down all the evolutionary dynamics, hence it takes longer for the fittest trait to be selected. Furthermore, in this case $\bar{y}=1$, meaning that the theoretical results need to take into account the fact that the fittest trait is not in the interior of $Y$; the convergence to $1$ necessarily takes place from below and this makes the uninfected population more susceptible to the infection. We remark that longer numerical simulations confirm the convergence towards $1$ with the associated cell densities (not shown). However, the equilibrium is reached only after a very long time, therefore from the application point of view we should note that the treatment outcome is slightly better than expected (although still not successful).

\subsubsection{\giulia{Inhomogeneous oxygen distribution (\exponeb)}}
\label{subsubsec:case1b}
\begin{figure}
	\centering
	\includegraphics[width=0.75\linewidth]{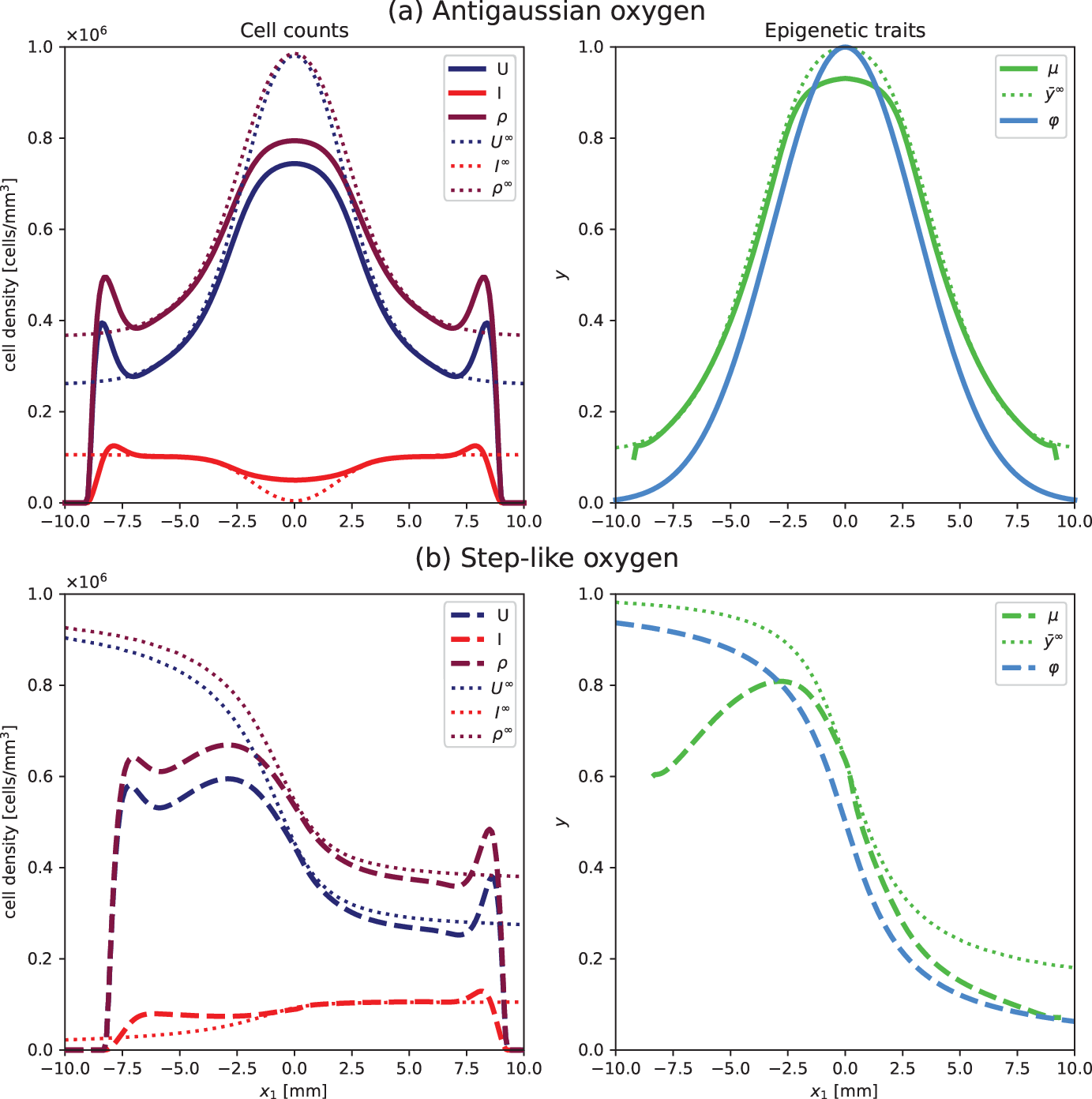}
	\caption{Results of the numerical simulations for stationary oxygen with antigaussian and step-like profiles at time $t^*$ corresponding to approximately $1800\;$h after viral injection. Only the horizontal section is shown to facilitate the comparison with the equilibria. All the graphical elements have the same meaning as in Fig. \ref{fig:time_evolution}.}
	\label{fig:hom_stat_antigaussian}
\end{figure}


\begin{figure}
	\centering
	\includegraphics[width=0.75\linewidth]{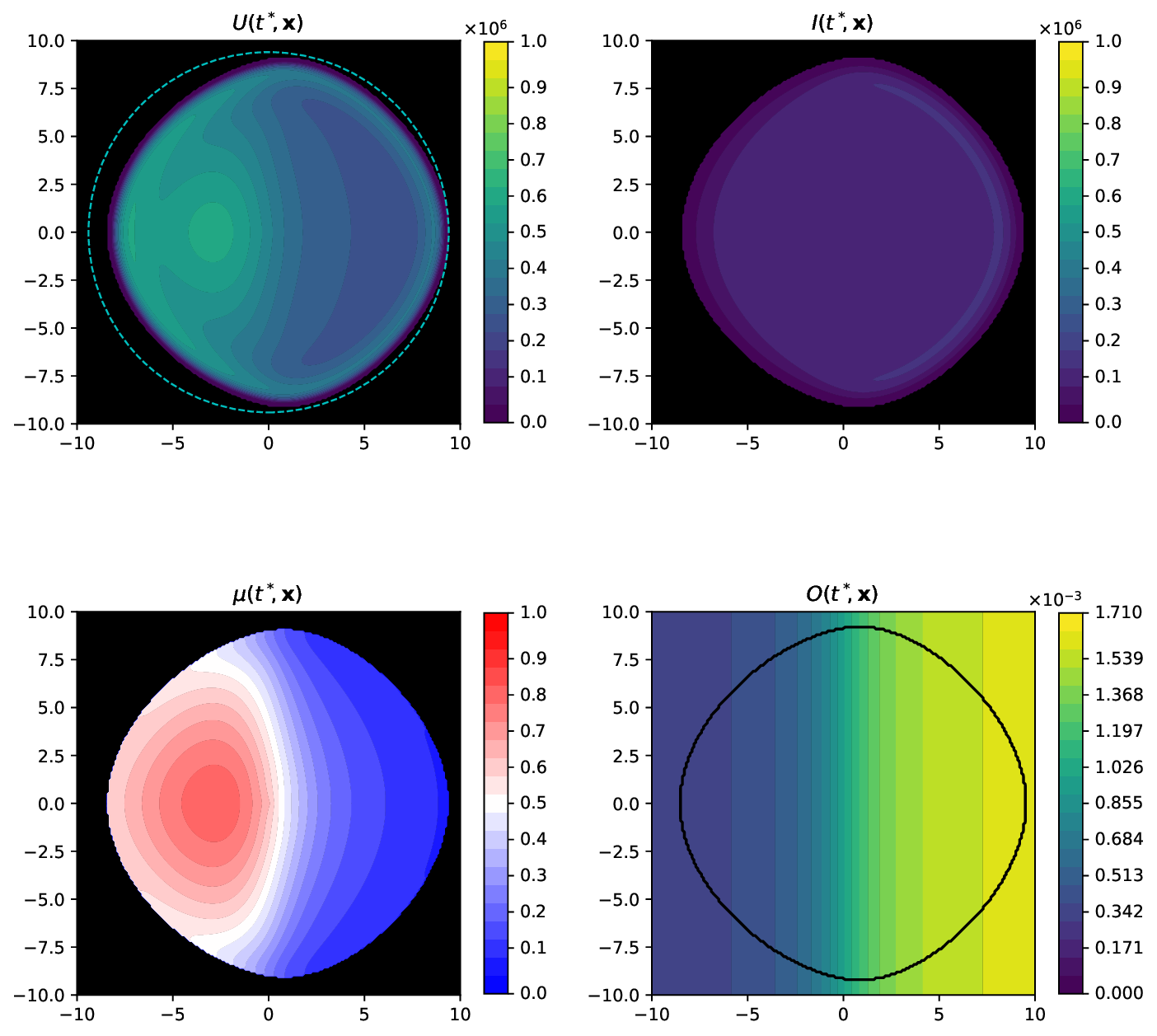}
	\caption{Results of the numerical simulations for stationary oxygen with step-like profile at time $t^*$ corresponding to approximately $1800\;$h after viral injection. We now show the densities in the whole domain to highlight the anisotropy. The circle in the plot of $U(t^*,\vec{x})$ is centred in $(0,0)$ and the radius is the distance from the origin of the furthest point $\vec{x}$ such that $\rho(t^*,\vec{x})>0$. On the other hand, the dark line in the plot of $O(t^*,\vec{x})$ encloses the region in which $\rho(t^*,\vec{x})>0$.}
	\label{fig:hom_stat_arctan_c}
\end{figure}

We now increase the model complexity by considering spatial homogeneities of the oxygen. We consider the following oxygen spatial profiles:
\begin{enumerate}
	\item ${O}(\vec{x}) = O_m+(O_M-O_m) e^{\frac{\abs{\vec{x}}^2}{20}} $ (antigaussian);
	\item ${O}(\vec{x}) = O_m+\Bigl[\dfrac{1}{2}+\dfrac{1}{\pi}\,\arctan\Bigl(\dfrac{x_1}{2}\Bigr)\Bigr] (O_M-O_m)$ (step-like profile).
\end{enumerate} 
The first profile is chosen to qualitatively resemble the oxygen distribution obtained after the tumour's consumption and we expect to observe dynamics somehow similar to the ones described below when we drop the stationarity assumption. The second one represents a tumour that grows at the boundary between two regions with significantly different vascularisation.

The numerical results of these two cases are shown in Fig. \ref{fig:hom_stat_antigaussian}, along with the videos accompanying it (see electronic supplementary materials S3 and S4). We again represent the solutions approximately $1800\;$h after the viral injections, which take place respectively at times $t=565\;$h and $t=470\;$h. The extremely slow growth in the first situation is due to the severely hypoxic conditions that characterise the initial growth: although this is probably unrealistic, later dynamics appear comparable to biologically meaningful scenarios. In both cases, we observe behaviours coherent with the previous findings, with the invasion led by slightly more proliferative cells and the slower convergence to the theoretical equilibrium in the hypoxic areas. The significance of these situations is the emergence of new selective dynamics that occur when the tumour reaches areas with different oxygen levels: in this respect, the effectiveness of virotherapy differs significantly from point to point. 

Another significant aspect is the nonsymmetrical configuration of the step-like profile, which allows to analyse the influence of the oxygenation on the front speed. Fig. \ref{fig:hom_stat_arctan_c} shows the result of the simulation in the whole domain and it is clear that the average epigenetic trait at the front significantly differs in the different directions. We recall that, in the case of pressure-driven movement, the \david{theoretical} front speed is higher for high cell proliferation rate and high cell density. In the case of our interest, fast proliferation is associated with effective viral infection, which results in a lower cell density: as a consequence, a priori, it is not trivial to understand which conditions are associated with a faster mass growth. The circle in Fig. \ref{fig:hom_stat_arctan_c} elucidates this aspect well, by showing that the fastest invasion still occurs in the most oxygenated area.

\subsection{\giulia{The dynamic oxygen case: comparing standard and hypoxia-specific oncolytic viruses in a full scenario}}
\label{subsec:dyn}


\giulia{After making initial observations on the effects of different oxygen concentrations on tumour evolution and therapy efficacy, it is natural to ask how the tumour dynamics co-depends on temporal variation of oxygen levels. We therefore consider the whole dynamics of Eq. \eqref{eq:complete}, in which oxygen varies both in space and time according to Eq. \eqref{eq_O}. At this level of complexity, which closely approaches an “in vivo” scenario compared to the earlier experiments, the metabolic processes of the tumour mass influence environmental conditions. On the other hand, the full dynamics also result into a slower convergence towards the asymptotic equilibria predicted by the theoretical analysis. We explore both the kind of viruses considered in the previous sections (see Section \ref{subsubsec:case2a}) and an hypoxia-specific oncolytic virus (see Section \ref{subsubsec:case2b}).

}

We consider a source of the form 
\[
Q(\vec{x}) = q_O \bar{O}(\vec{x})
\]
where $\bar{O}(\vec{x})$ is the oxygen profile that we would observe in the absence of the tumour. We remark that the actual oxygen distribution is always below these values due to the increased oxygen consumption of cancer cells. 

\begin{figure}
	\centering
	\includegraphics[width=0.75\linewidth]{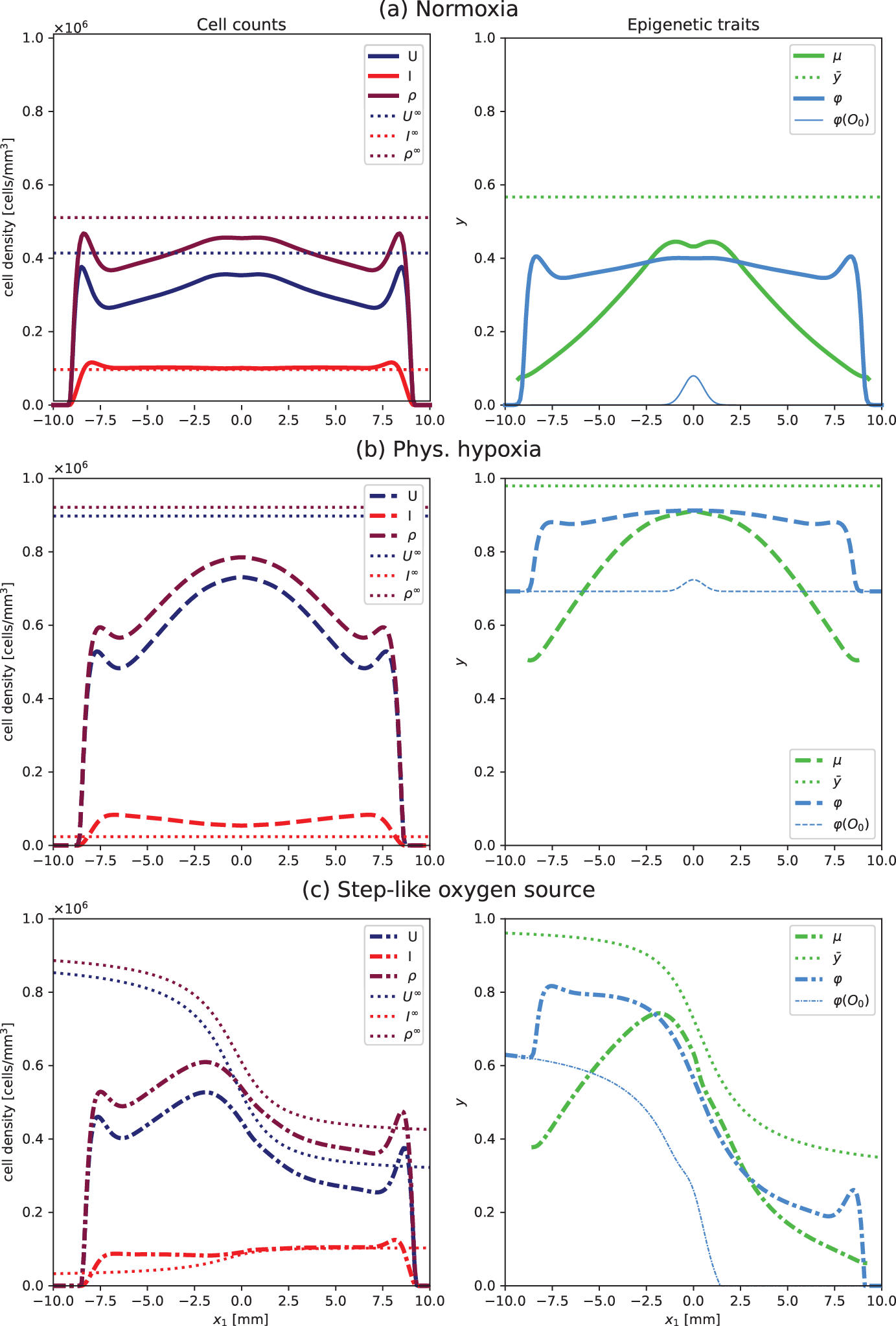}
	\caption{Comparison of the results obtained from numerical simulation of the full model at time $t^*$, corresponding to approximately $1800\;$h after viral injection. Three spatial oxygen conditions are considered: normoxia (solid lines, panel (a)), hypoxia (dashed lines, panel (b)), and step oxygen source (dot-dashed lines, panel (c)). For explicit formulation, see the main text. All the graphical elements have the same meaning as in Fig. \ref{fig:time_evolution}.In addition, a thinner light blue line in the right panels provides the $\varphi$ profile at the initial state. The equilibria are now computed from Eq. \eqref{eq:equilibria_complete}.} 
	\label{fig:dyn_normox}
\end{figure}

We use as initial condition $O_0(\vec{x})$ the steady state of Eq.~\eqref{eq_O}, i.e., the solution of the equation
\begin{equation*}
	{D_O \Delta O(t,\vec{x})} - {q_O O(t,\vec{x})}  - {\lambda \rho(0,\vec{x}) O(t,\vec{x})} + {Q(\vec{x})} = 0
\end{equation*}
where $\rho(0,\vec{x}) = \int_{\Omega} u_0(x,y)\di y$ and $u_0$ is given in Eq. \eqref{eq:u_init}. We remark that our parameter choice allows to have the same values of $\rho(0,\vec{x})$ for all $y_0$; as a consequence, it still makes sense to define $y_0=\varphi(O_0(\vec{0}))$.

\subsubsection{\giulia{Standard oncolytic viruses (\exptwoa)}}
\label{subsubsec:case2a}

\david{We consider the effect of standard oncolytic virotherapy in the following oxygen profiles:
\begin{enumerate}
	\item $\bar{O}(\vec{x}) = O_M $ (normoxia);
	\item $\bar{O}(\vec{x}) = 3\,O_m $ (hypoxia);
	\item $\bar{O}(\vec{x}) = 3\,O_m+\Bigl[\dfrac{1}{2}+\dfrac{1}{\pi}\,\arctan\Bigl(\dfrac{x_1}{2}\Bigr)\Bigr] (O_{\max}-3\, O_m)$ (step-like source).
	\item  $\displaystyle \bar{O}(\vec{x}) = 3\,O_m+(O_M-3\, O_m)\sum_{k=1}^{3} e^{-\frac{|\vec{x}-\vec{x}_k|^2}{15}}$ with $\vec{x}_1\coloneqq (-4,4)$, $\vec{x}_2\coloneqq (3,-6)$ and $\vec{x}_3\coloneqq (6,-3)$ (source with three peaks).
\end{enumerate} 
The homogeneous sources aim at reproducing uniformly vascularised tissues, which, in the absence of a tumour, are either normoxic or physiologically hypoxic. The step-like source models a tissue with two distinguishable areas due to different oxygen inflow rates. Finally, the last source profile constitutes an example of a tissue in which heterogeneous vascularisation leads to a varied oxygen profile. A spatially heterogeneous oxygen source allows us to consider vessels of different sizes and, thus, with variable blood flow.}

\begin{figure}
	\centering
	\includegraphics[width=0.75\linewidth]{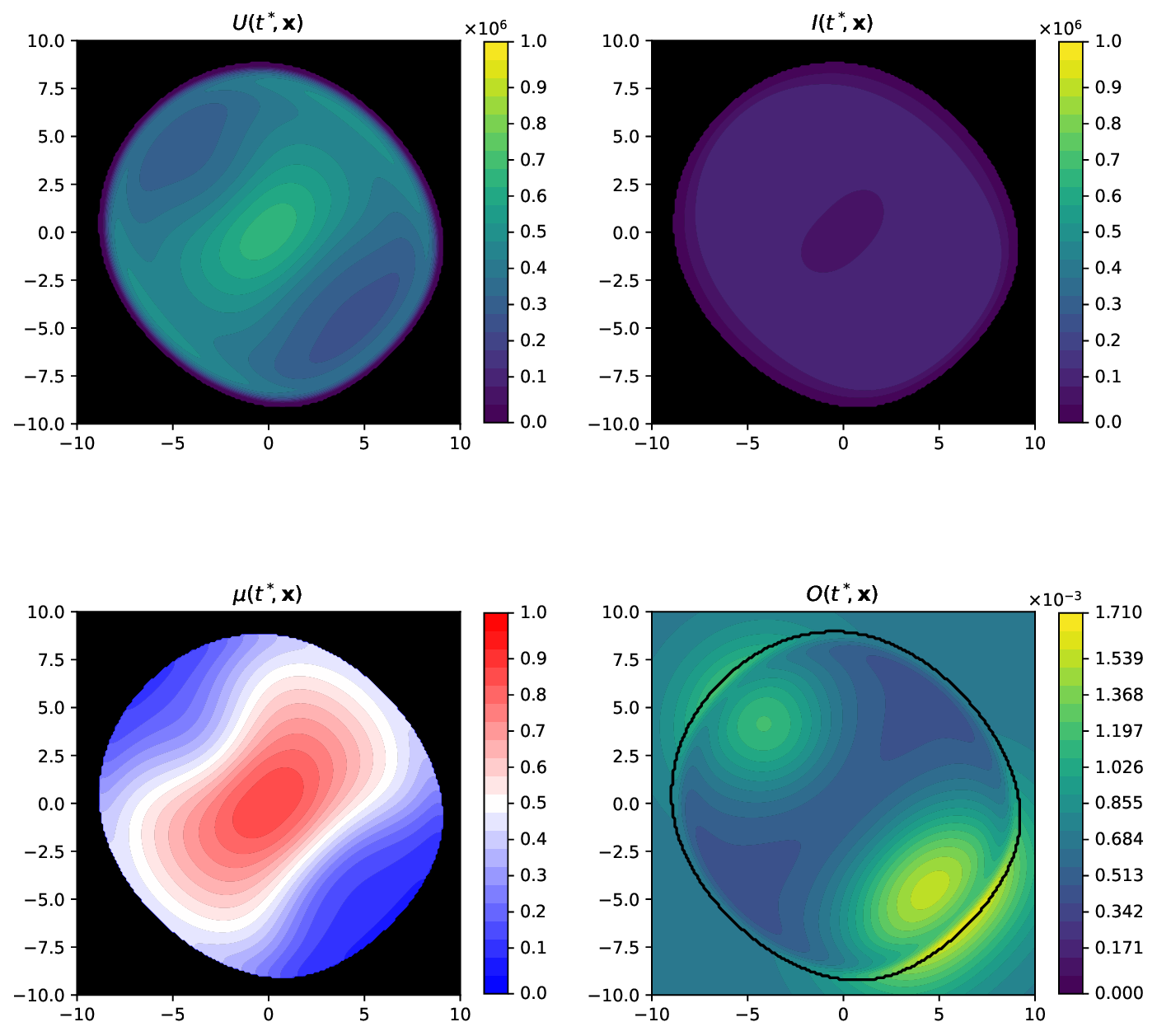}
	\caption{Results of the numerical simulation for the full model with oxygen source with three peaks at time $t^*$ corresponding to approximately $1800\;$h after viral injection. We now show the densities in the whole domain to highlight the anisotropy. The dark line in the plot of $O(t^*,\vec{x})$ encloses the region in which $\rho(t^*,\vec{x})>0$.}
	\label{fig:multi}
\end{figure}

The results are collected in Figs. \ref{fig:dyn_normox} and \ref{fig:multi}, along with the videos accompanying them (see electronic supplementary materials S5, S6 and S7),. The arrangement of plots in Fig. \ref{fig:dyn_normox} is analogous to Fig. \ref{fig:hom_stat_comparison}, with the inclusion of $\varphi(O(0,\vec{x}))$ in the right panels: this allows to quantify the variation in time of the oxygen concentration due to tumour growth and the consequent evolution of the trait selected by environmental conditions. Fig. \ref{fig:multi} shows the simulation result in the whole domain and highlights how the source heterogeneity affects the dynamics.

In all four cases, the initial tumour growth causes a drop in oxygen concentration, reducing the environmentally optimal epigenetic trait; consequently, the tumour growth progressively slows down. The variation of oxygen level is a new selective pressure, which could not be considered in the stationary oxygen situation. The dynamics in the centre of the tumour are characterized by a progressive adaptation, with the oxygen that reduces as the cell density grows and the average epigenetic trait that increases as the oxygen density decreases. We remark that in this initial phase, the actual fittest trait is always lower than the one which is optimal for the oxygen-driven selective pressure due to the distance from carrying capacity (see Eq. \eqref{eq:ybar}); as time passes, this difference becomes less evident. In the meantime, the invasion fronts are more oxygenated, which contributes to the selection of proliferative traits in this area; this behaviour resembles the dynamics observed in the case of antigaussian stationary oxygen distribution. Spatial heterogeneity affects the front behaviour in the cases of step-like and multiple peak sources.

As with stationary oxygen, the viral injection constitutes an additional selective pressure. The virotherapy causes a significant decrease in the cancer population, which allows the reoxygenation of the tissue. Nevertheless, the initial selective pressure of the infection appears more substantial than the environmental pressure, and the average epigenetic trait significantly increases. As time passes, the infection is more effective in the well-oxygenated areas: this keeps the cancer population low and avoids an excessive reduction in the oxygen concentration, which would result in a less effective infection. Conversely, in the less oxygenated areas, the tumour grows up to close to carrying capacity; hence, the oxygen concentration reduces further and the environmental conditions contribute to the selection of cells resistant to the infection.

Fig. \ref{fig:dyn_normox} shows that the solutions of the equation approach the theoretical estimates, but the convergence is slower than in the case of stationary oxygen. This is particularly evident in the hypoxic areas due to the slow evolutionary dynamics related to the low growth rate, as already observed for stationary oxygen, and it is now accentuated by the fact that it takes time for the oxygen distribution to reach equilibrium. Furthermore, our tools do not allow to characterise the wavefront, whose dynamics are significantly far from equilibrium. Nevertheless, the theoretical values still provide significant information regarding the success of the therapy. 

Overall, the main dynamics observed in Section \ref{subsec:stat} can be replicated without fixing a priori the oxygen distribution (which is not representative of realistic biological scenarios); oxygen dynamics significantly enriches the evolutionary dynamics. Initial condition referable as normoxic (panel (a)) or physiologically hypoxic (panel (b)) in the absence of the tumour become respectively physiologically and severely hypoxic due to cancer: in this sense, these two settings can be considered as a ``trait d'union'' with the previous simulations. Similar dynamics can also be observed in the presence of more complex oxygen sources, such as the three-peaked source of Fig. \ref{fig:multi}: this suggests that the knowledge of the oxygen distribution in a tumour may predict the outcome of the virotherapy in clinical settings.

\subsubsection{\giulia{Hypoxia-specific oncolytic viruses (\exptwob)}}
\label{subsubsec:case2b}

All the simulations presented so far rely on the assumption that viral infection is less effective in hypoxic cells: this can be explained by their slower metabolic activity, which also affects the translation of viral proteins \cite{shengguo11}. Nevertheless, one should note that some particular oncolytic viruses can specifically target receptors that are upregulated in case of the lack of oxygen \cite{sadri23,shengguo11}. This property appears particularly promising in light of the ineffectiveness of most classic cancer therapies in hypoxic conditions \cite{zhuang23}. 

Therefore, we revert the previous trade-off and exchange the values of $\beta_M$ and $\beta_m$, so that the function $\beta(y)$ is increasing; the rest of the model remains unchanged. The asymptotic analysis of \ref{app:theory} does not rely on any characterisation of the values of $\beta$ and all the equations obtained are still valid: the only relevant difference is the fact that Eq. \eqref{eq:ybar} now predicts a decrease of the fittest value in the presence of viral infection. The equilibrium values depicted in 
Fig. \ref{fig:equilibria}b indeed shows that the virotherapy's effectiveness increases as the oxygen concentration decreases, in line with the biological situation we aim to model.

\begin{figure}
	\centering
	\includegraphics[width=0.75\linewidth]{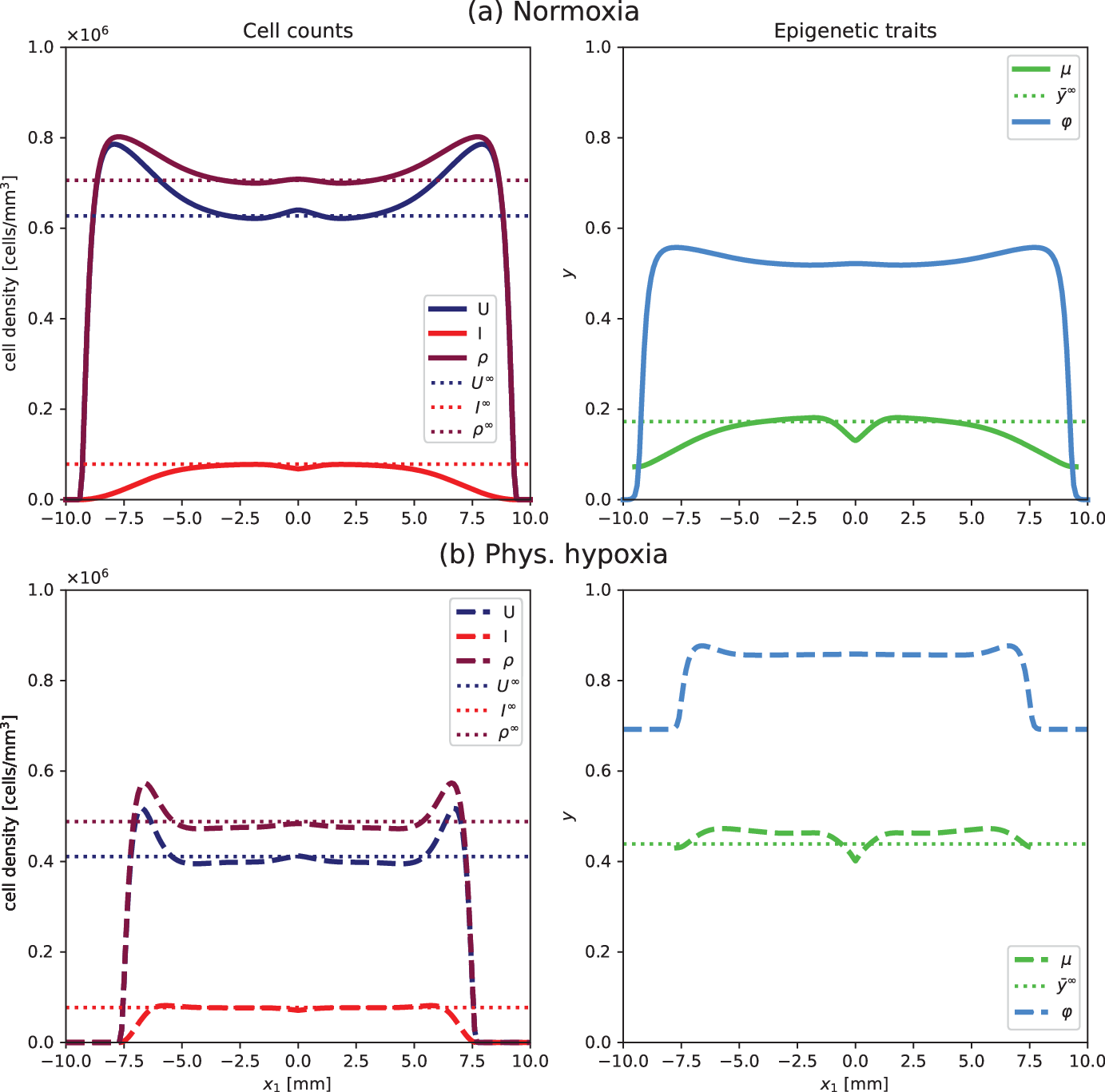}
	\caption{Comparison of the results obtained from numerical simulation of the full model in the case of hypoxia-targeting oncolytic viruses at time $t^*$, corresponding to approximately $1500\;$h after viral injection. We now consider an earlier time than in the previous simulations, as the infection in this case promotes the selection of more proliferative cells and, at later times, we would observe boundary effects in the normoxic case. Two spatial oxygen conditions are considered: normoxia (solid lines, panel (a)) and hypoxia (dashed lines, panel (b)); the oxygen sources are the same as in Figs. \ref{fig:dyn_normox}a-b. All the graphical elements have the same meaning as in Fig. \ref{fig:time_evolution}.}
	\label{fig:inverse}
\end{figure}

Fig. \ref{fig:inverse}, along with the video accompanying it (see electronic supplementary material S8), shows the full model results for the oxygen sources corresponding to normoxia and hypoxia. In the case of normoxia (Fig. \ref{fig:inverse}a), we observe the failure of the therapy: despite the persistence of the infection, the uninfected cell density is now approximately twice the value observed in Fig. \ref{fig:dyn_normox}a; furthermore, the expansion of the tumour is much faster than in the case of virotherapy with standard viruses that target non-hypoxic cells \david{(in line with the higher theoretical wave speed associated to lower values of $y$)}. Both these phenomena are caused by the fact that viral infection selects cells with a value of the epigenetic trait lower than the one which is optimal for the oxygen-driven selective pressure; these cells are both more proliferative and more resistant to the infection. Interestingly, the results are not trivial: indeed, a priori, one could expect that the hypoxic condition caused by the tumour growth would be associated with a large number of hypoxic cells, which are effectively targeted by the virus. Our mathematical model is thus helpful to shed light on these complex interactions.

The situation of physiological hypoxia appears much more promising (Fig. \ref{fig:inverse}b). Again, we observe a decrease in the average epigenetic trait caused by virotherapy, as expected from the theoretical results; consequently, the front speed of the tumour is also higher. Nevertheless, the cell lines selected are not too resistant to the infection and the tumour cell density appears comparable to that observed in the case of normoxia treated by a standard virus. The important consequence is that the effectiveness of oncolytic virotherapy can be increased by selecting the most appropriate kind of virus based on the oxygenation of the tumour.

\begin{figure}
	\centering
	\includegraphics[width=0.75\linewidth]{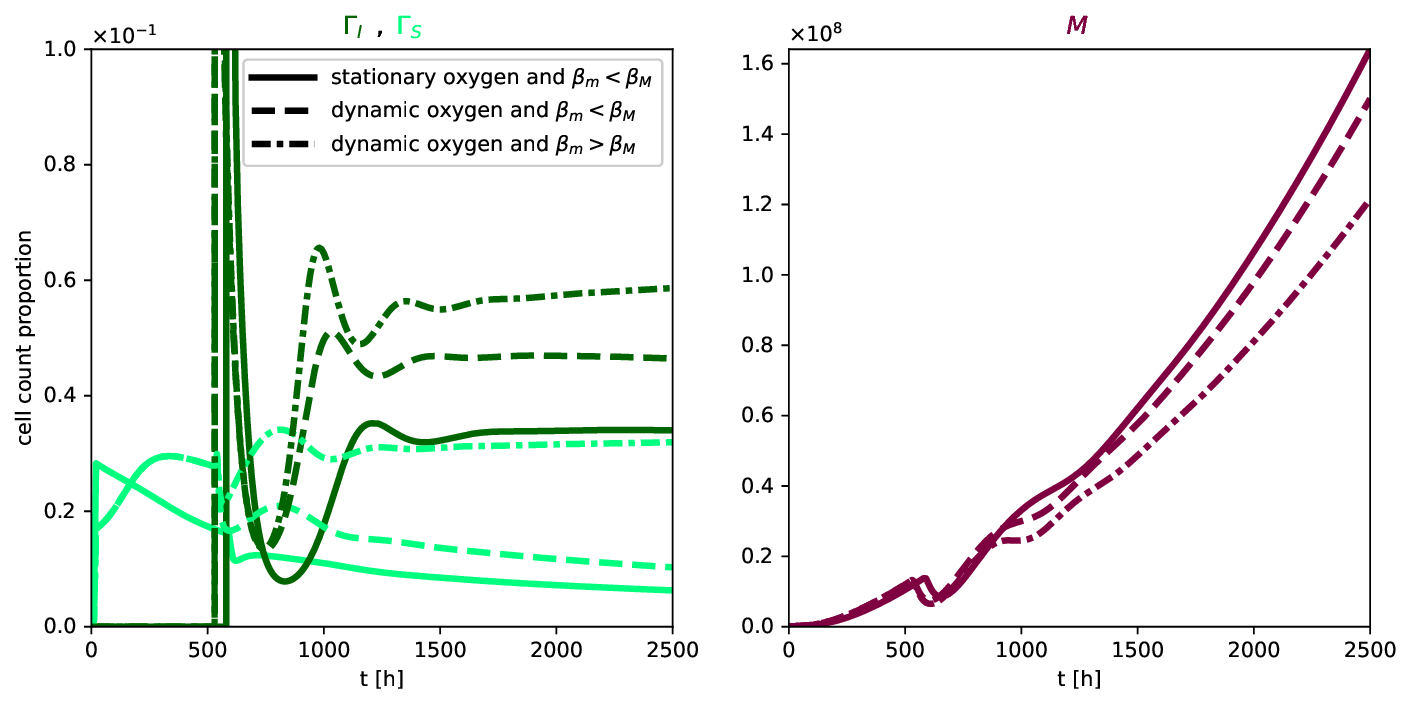}
	\caption{Comparison of the time evolution of the proportion of cells killed by environmental-driven selective pressure $\Gamma_I$, the proportion of tumour cells infected at a given instant  $\Gamma_S$ and the total number of cancer cells $M$ in the cases of stationary oxygen and $\beta_m<\beta_M$ (solid lines), dynamic oxygen and $\beta_m<\beta_M$ (dashed lines), dynamic oxygen and $\beta_m>\beta_M$ (dot-dashed lines). In the case of stationary oxygen, the oxygen concentration is the equilibrium value of the corresponding model with dynamic oxygen (as depicted in the right panel of Fig. \ref{fig:dyn_normox}b).}
	\label{fig:hyp_comparison}
\end{figure}

The comparison between the two kinds of oncolytic viruses in the case of physiological hypoxia is further elucidated in Fig. \ref{fig:hyp_comparison}, in which we consider the total number of cancer cells in the domain
\[
M(t) \coloneqq \int_{\Omega} \rho(t,\vec{x})\di\vec{x}.
\]
We also consider the evolution in time of the proportion of cells killed by environmental-driven selective pressure and the proportion of tumour cells infected at a given instant, respectively 
\begin{gather*}
\Gamma_S(t)\coloneqq \frac{\eta}{M(t)}\int_Y\int_{\Omega} (y-\varphi(O(t,\vec{x})))^2 u(t,\vec{x},y)\di\vec{x}\di y, \\
\Gamma_I(t)\coloneqq \frac{1}{M(t)}\int_Y\int_{\Omega} \beta(y) u(t,\vec{x},y) v(t,\vec{x})\di\vec{x}\di y.
\end{gather*}
In all cases, the main contribution to cell death is caused by viral infection, meaning that the therapy is always at least partially effective. The solid lines refer to the situation in which the oxygen concentration is not affected by the tumour and allow us to understand the role of oxygen dynamics; for the sake of brevity, this simulation is only performed in the case $\beta_m<\beta_M$. The comparison with the analogous simulation of the full model (dashed lines) shows that more cells are killed by environmental-driven selective pressure in the case of variable oxygen: indeed, hypoxia exerts a more significant action when the tumour needs to adapt to an evolving environment. The viral infection is more efficient in the latter case since it takes a long time for the tumour to fully adapt to the hypoxic state and become resistant to the infection. When the hypoxia-specific virus is considered, the infection appears significantly more effective. It is also remarkable that the environmental-driven selective pressure is more relevant in the last situation: indeed, the value of $\bar{y}$ given by Eq. \eqref{eq:ybar} is now much lower than $\varphi(O)$ (see again Figs. \ref{fig:dyn_normox}b and \ref{fig:inverse}b), as for $\beta_M<\beta_m$ both terms in the square bracket are lower or equal to $0$; in other words, the infection favours cells that are less resistant to hypoxia. We should therefore conclude that the therapeutic effectiveness is also due to environmental-driven selective pressure. 

Although the use of hypoxia-selective oncolytic viruses appears promising, our findings also stress that tumour eradication is still far from achieved. Furthermore, the cell lines selected are the most proliferative; hence, if the viral infection stopped for some external reason (such as immune response), the tumour would quickly regain its aggressiveness. Nonetheless, it is essential to observe that the average epigenetic trait is significantly reduced with respect to the one that environmental conditions would select; consequently, the tumour is now sensitive to standard treatments that lose their effectiveness in hypoxic conditions. This suggests that a hypoxic tumour could be effectively treated using a combination of therapies targeting cells with different degrees of adaptation to hypoxia \cite{shengguo11}.

\section{Conclusions}
\label{sec:conclusion}

In this work, we have introduced a novel oncolytic virotherapy model that considers the epigenetic evolution of cancer cells due to viral infection and hypoxia. By integrating viral dynamics, tumour evolution, and spatial oxygen gradients, our model provides new insights into how hypoxic conditions within tumours affect the efficacy of oncolytic viral therapies. Numerical simulations are coherent with the theoretical results obtained by a formal asymptotic analysis of simplified settings and show how environmental conditions influence the capability of the virus to control tumour mass expansions and, in particular, underscore the significant impact of oxygen availability on viral infection rates, tumour cell susceptibility and the overall success of virotherapy. We have considered some simple configurations of oxygen sources to capture the fundamental dynamics. Our results suggest that hypoxia may constitute a significant obstacle to the success of oncolytic virotherapy. Furthermore, the infection contributes to selecting a subpopulation of cells well adapted to hypoxia, which may be hard to kill even with other therapies. Hypoxic tumours appear to be more effectively treated with oncolytic viruses specifically targeting hypoxic cells, although tumour eradication still appears hard to achieve. 

Our findings highlight several crucial aspects for optimising virotherapy: the presence of hypoxic regions can severely limit the spread and replication of standard oncolytic viruses; in contrast, hypoxia-specific viruses are particularly effective in these areas. This suggests that therapeutic strategies need to account for spatial oxygen level variations. The evolutionary dynamics of tumour cells under hypoxic stress and virotherapy pressure can lead to the emergence of resistant phenotypes, indicating the need for adaptive treatment protocols that can mitigate resistance development. One critical takeaway is the importance of considering the spatial heterogeneity of oxygen levels when designing and implementing oncolytic virotherapy protocols. The model predicts that hypoxia can significantly alter the distribution and effectiveness of viral therapy, thereby affecting overall treatment outcomes. Additionally, the role of tumour cell adaptations to hypoxic conditions highlights the necessity for dynamic treatment strategies that can respond to changes in the tumour microenvironment.

The key assumption of this work is that, in general, the slower metabolic activity of hypoxic cells is associated with a less effective viral infection. This has been modelled by considering an infection rate $\beta$ that depends on the adaptation of cells to hypoxia. Our choice is motivated by the existing literature \cite{lorenzi21} and the possibility of obtaining some theoretical insights. Other approaches would also be possible: for example, one may directly consider the dependence of viruses on the host translational machinery to translate viral proteins; this would correspond to a viral burst size $\alpha$ that depends on the oxygen concentration or on the level of metabolic activity of the cell (which could correspond to the variable $y$).

There are several promising directions for future research and the results may be extended in several ways, both from the mathematical and the modeling perspectives. From the mathematical point of view, the formal asymptotic analysis of \ref{app:theory} may constitute a good starting point for characterising the travelling waves shown in the numerical simulations, at least in simplified settings. From a biological point of view, we aim to develop a model to study the interaction of oncolytic virotherapy and radiotherapy, which aligns with the clinical interest of combining the two therapies. Radiotherapy is well-known to decrease its effectiveness in hypoxic conditions and several mathematical models similar to the one in the present work have been developed to investigate this phenomenon \cite{celora21,chiari23radio}. Although the combination with oncolytic viruses that decrease their efficacy due to hypoxia does not appear beneficial, using hypoxia-targeting oncolytic viruses could be promising.

Overall, our work contributes to a deeper understanding of the complexities of oncolytic virotherapy under hypoxic conditions and lays the groundwork for developing more effective and personalised cancer treatments. In this context, mathematical models could help design the optimal combination, considering contemporary, subsequent, or alternating treatments and investigating doses, orders and timing according to the environmental conditions.

\appendix

\section{Formal asymptotic analysis}
\label{app:theory}

We now conduct a formal asymptotic analysis to compute the theoretical equilibrium values in the spatially homogeneous case, as explained in Section \ref{sec:theory}. For the sake of simplicity, we neglect the space variable $\vec{x}$. We introduce a small parameter $\epsilon$ and assume that $D_y=\epsilon^2$. In light of the time scaling $t\mapsto \frac{t}{\epsilon}$, we define 
\begin{gather*}
u_\epsilon(t,y)\coloneqq u\Bigl(\frac{t}{\epsilon},y\Bigr), \qquad I_\epsilon(t)\coloneqq I\Bigl(\frac{t}{\epsilon}\Bigr), \\ v_\epsilon(t)\coloneqq v\Bigl(\frac{t}{\epsilon}\Bigr), \qquad O_\epsilon(t)\coloneqq O\Bigl(\frac{t}{\epsilon}\Bigr).
\end{gather*}
The system of Eq. \eqref{eq:complete} then becomes
\begin{equation}
	\label{eq:scaled_hyp}
	\begin{cases}
		\epsilon\pt u_\epsilon(t,y)= \epsilon^2 \pyy u_\epsilon(t,y)\\
		\phantom{\epsilon\pt u_\epsilon(t,y)=}+ R(y,\rho_\epsilon(t),O_\epsilon(t),v_\epsilon(t))\; u_\epsilon(t,y)\\
		\epsilon\pt I_\epsilon(t)= v_\epsilon(t) \int_Y \beta(y) u(t,y) \di y\\
		\phantom{\epsilon\pt I_\epsilon(t)=}-q_I I(t) \\
		\epsilon\pt v_\epsilon(t)= \alpha q_I I_\epsilon(t) -q_v v_\epsilon(t) \\
		\epsilon\pt O_\epsilon(t)= -q_O O_\epsilon(t) -\lambda \rho_\epsilon(t) O_\epsilon(t) + Q\\
		\rho_\epsilon(t)\coloneqq \int_Y u_\epsilon(t,y)\di y +I_\epsilon(t)
	\end{cases}
\end{equation}
Let us observe that, letting $\epsilon\to 0$ and assuming that all the functions converge, we immediately get from the third equation that
\begin{equation}
	\label{eq:v}
	v(t)= \frac{\alpha q_I}{q_v} \, I(t)
\end{equation}
and from the second equation that
\begin{equation}
	\label{eq:I_hyp}
	I(t)=0 \quad \text{or} \quad  \int_Y \beta(y) u(t,y)\di y= \frac{q_I I(t)}{v(t)}=\frac{q_v}{\alpha}.
\end{equation}
Furthermore, the fourth equation yields
\begin{equation}
	\label{eq:O}
	O(t)=\frac{Q}{q_O+\lambda\rho(t)}.
\end{equation}
It is important to note that the system may not converge to an equilibrium: indeed, for some parameter values, central oscillations persist even for very long times, similar to the ones described in Refs. \cite{baabdulla23,pooladvand21} (see also electronic supplementary material S9).

As explained in Section \ref{sec:theory}, we make the ansatz
\[
u_\epsilon(t,y)= e^{\frac{n_\epsilon(t,y)}{\epsilon}}.
\]
This implies
\[
\pt u_\epsilon =\frac{\pt n_\epsilon}{\epsilon}u_\epsilon, \quad \pyy u_\epsilon = \Bigl(\frac{(\py n_\epsilon)^2}{\epsilon^2}+\frac{\pyy n_\epsilon}{\epsilon}\Bigr) u_\epsilon.
\]
The first equation of Eq. \eqref{eq:scaled_hyp} yields
\[
\pt n_\epsilon = (\py n_\epsilon)^2 +\epsilon\pyy n_\epsilon+ R(y,\rho_\epsilon,O_\epsilon,v_\epsilon).
\]
Letting $\epsilon\to 0$ and assuming convergence, we obtain
\begin{equation}
	\label{eq:n_hyp}
	\pt n = (\py n)^2 +\ R(y,\rho,O,v).
\end{equation}
All the functions without the subscript $\epsilon$ are the leading-order terms of the asymptotic expansion.

Under the concavity hypotheses explained in Section \ref{sec:theory}, we expect $n_\epsilon$ to be a strictly concave function of $y$; this is also true for $n$ \cite{mirrahimi2015asymptotic, perthame08}. We then define
\[
\bar{y}(t)\coloneqq \arg\max_{y\in Y} n(t,y).
\]
It is clear from the definition of $R$ that
\[
\partial_O R \leq 0, \qquad \partial_v R\leq 0, \qquad R(y,0,0,0)=R(y,K,0,0)=0 \quad \forall y\in Y.
\]
As a consequence, unlimited growth of cancer cells is not possible, i.e. $\rho_\epsilon(t)<+\infty$ for all $\epsilon$ (see also, for example, Ref. \cite{lorenzi21invasion}). Therefore,
\[
n(t,\bar{y}(t))=\max_{y\in Y} n(t,y)=0
\]
and whenever $\bar{y}\in (0,1)$ trivially
\[
\py n(t,\bar{y}(t))=0.
\]
In the rest of the analyses we assume $\bar{y}\in (0,1)$ and refer to the main text for the cases $\bar{y}\in \{0,1\}$. We also observe that
\[
0=\frac{\partial}{\partial t}  n(t,\bar{y}(t))= \pt  n(t,y)|_{y=\bar{y}(t)} + {\py n(t,y)|_{y=\bar{y}(t)}} \pt \bar{y}(t)
\]
implying that $\pt n(t,y)|_{y=\bar{y}(t)}=0$.

We evaluate Eq. \eqref{eq:n_hyp} in $y=\bar{y}(t)$ to get
\begin{equation}
	\label{eq:R_hyp}
	\begin{split}
		[p_M+(p_m-p_M)y] \Bigl( 1-\frac{\rho}{K}\Bigr) - \eta (y-\varphi(O))^2 -[\beta_M+(\beta_m-\beta_M)\bar{y}(t)] v=\\
		=R(\bar{y}(t),\rho(t),O(t),v(t)) =\pt n - (\py n)^2=0.
	\end{split}
\end{equation}
We can also differentiate Eq. \eqref{eq:n_hyp} with respect to $y$ to get
\[
\partial_{ty}^2 n = 2\py n\, \pyy n +\py R(y,\rho,O,v) 
\]
which computed at $y=\bar{y}(t)$ yields
\[
\partial_{ty}^2 n(t,\bar{y}(t))= \py R(\bar{y}(t),\rho(t),O(t),v(t)) 
\]
Let us now look for a steady state, which we denote by $(U^\infty,I^\infty,v^\infty,O^\infty,\bar{y}^\infty)$. 
The previous equation implies
\begin{equation}
	\label{eq:derR}
	\py R(\bar{y}^\infty,\rho^\infty,O^\infty,v^\infty)=(p_m-p_M)\Bigl( 1-\frac{\rho^\infty}{K}\Bigr) - 2\eta (\bar{y}^\infty-\varphi(O^\infty)) -(\beta_m-\beta_M) v^\infty=0.
\end{equation}
It is reasonable to consider $u_\epsilon\overset{*}{\rightharpoonup} U \,\delta_{\bar{y}}$, in the meaning of weak-$*$ convergence of measures, i.e.
\[
\int_Y \psi(y) u_\epsilon(t,y) \di y\xrightarrow{\epsilon\to 0} \psi(\bar{y}(t)) U(t) \quad \forall \psi\in C^0(Y) .
\]
We remark that the convergence only involves the epigenetic variable and is not affected by the addition of space: indeed, in the general case we can rely on Eq. \eqref{eq:selectedt} for all $\vec{x}\in\supp(\rho)$. In the spatially homogeneous case, we obtain
\[
\int_Y \beta(y) u_\epsilon(t,y)\di y \to \beta(\bar{y}(t)) {U}(t).
\]
With this observation, Eqs. \eqref{eq:v}, \eqref{eq:I_hyp}, \eqref{eq:O}, \eqref{eq:R_hyp} and \eqref{eq:derR} constitute a system of five equations in the five variables $U^\infty,I^\infty,v^\infty,O^\infty,\bar{y}^\infty$, which in principle can be solved.

Let us first focus on the infection-free case of Eq. \eqref{eq:I_hyp}, i.e. $I^\infty=0$. This clearly implies $v^\infty=0$ and the other three variables solve the system
\begin{equation}
	\label{eq:eq_complete_noinf}
	\begin{cases}
		R(\bar{y}^\infty,U^\infty,O^\infty,0) =p(\bar{y}^\infty)\Bigl( 1-\dfrac{U^\infty}{K}\Bigr) - \eta (\bar{y^\infty}-\varphi(O^\infty))^2=0 \\[8pt]
		\py R(\bar{y}^\infty,U^\infty,O^\infty,0)=(p_m-p_M)\Bigl( 1-\dfrac{U^\infty}{K}\Bigr) - 2\eta (\bar{y}^\infty-\varphi(O^\infty))=0 \\[8pt]
		O^\infty=\dfrac{Q}{q_O+\lambda U^\infty}
	\end{cases}
\end{equation}
From the first two equations, we obtain 
\[
1-\frac{U^\infty}{K}=\frac{\eta (\bar{y}^\infty-\varphi(O^\infty))^2}{p(\bar{y}^\infty)}=\frac{2\eta(\bar{y}^\infty-\varphi(O^\infty))}{(p_m-p_M)}
\]
which admits two solutions: the first one is $U^\infty=K, \bar{y}^\infty=\varphi(O^\infty)$; the second one is
\begin{gather*}
	\bar{y}^\infty= \frac{2 p_M}{p_M-p_m}-\varphi(O^\infty)>\frac{2 p_M}{p_M-p_m}-1=\frac{p_M+p_m}{p_M-p_m}>1 \\ 
	U^\infty=K+\frac{4\eta p(\varphi(O^\infty))}{(p_m-p_M)^2}> K
\end{gather*}
but it clearly has no biological meaning. If we assume that the oxygen concentration is not affected by tumour dynamics, then we neglect the third equation of Eq. \eqref{eq:eq_complete_noinf} and the equilibria are the above ones. If we consider the full system, then the first equilibrium is given by
\begin{equation}
\label{eq:eq_K}
U^\infty=K, \qquad O^\infty=\frac{Q}{q_O+\lambda K},\qquad \bar{y}^\infty=\varphi(O^\infty).
\end{equation}
The second equilibrium could be obtained by computing the solutions of a second-degree equation in $O^\infty$; given the complexity of the expressions and the lack of biological meaning, we omit further details.

Let us now assume $I^\infty\neq 0$, which according to Eq. \eqref{eq:I_hyp} implies
\[
U^\infty=\frac{q_v}{\alpha \beta(\bar{y}^\infty)}.
\]
This leads to the system of Eq. \eqref{eq:equilibria_complete}, described in the main text.

\section{Details of numerical simulations}
\label{app:num}

\subsection{Parameter values}

Table \ref{tab:parameters} lists the parameters we adopt as a reference in the numerical simulations. The majority of the parameters has been estimated from the empirical literature, while a few others are specific to our formulation of the model and have been set to reasonable values in order to reproduce plausible dynamics. Our two-dimensional simulations represent the section of a tumour that is approximately homogeneous along the third spatial dimension (which can therefore be neglected) and parameters are estimated in a three-dimensional setting.

The maximal duplication rate of uninfected cells $p_M$, corresponding to the normoxic situation, has been taken equal to $\log(2)/24\;$h$^{-1}\approx 2.88 \times 10^{-2}\;$h$^{-1}$; the duplication time of 24 hours is among the fastest values reported in Ref. \cite{ke00} for glioblastoma. On the other hand, we assume that severely hypoxic cells duplicate in 48 hours, as done in Ref. \cite{martinez12}: this leads to a minimal proliferation rate $p_m=1.44 \times 10^{-2}\;$h$^{-1}$. The carrying capacity $K$ has been estimated assuming that a cell has diameter $10\;\mu$m$=10^{-2}\;$mm \cite[\S1.1]{lodish08}: this implies that the carrying capacity is $10^6\;$cells/mm$^3$.

The spatial diffusion coefficient of tumour cells $D_{\vec{x}}$ has been estimated from the experimental data of the U343 control group of  \cite{kim06}, as already done in Refs. \cite{morselli23,pooladvand21}. In these experiments, the tumour volume passes in 40 days from $70\;$mm$^3$ to $1000\;$mm$^3$, which corresponds to a change in the tumour radius from approximately $2.6\;$mm to approximately $6.2\;$mm. Hypoxia may play some role in the process, but this is not taken into account in their data: for the sake of simplicity, we assume a moderately hypoxic situation so that the proliferation rate takes the value $p(0.5)$ (i.e., the average between $p_M$ and $p_m$). Hence, the dynamics of uninfected cells in the absence of viral infection follow the equation
\[
\pt u(t,\vec{x})=D_{\vec{x}} \dive_{\vec{x}} (u(t,\vec{x}) \nabla u(t,\vec{x}))+ p(0.5)\Bigl(1-\frac{u(t,\vec{x})}{K}\Bigr) u(t,\vec{x}).
\]
As we mention in the main text, an initial condition with compact support evolves into a wave that travels with the minimal speed $\sqrt{D_{\vec{x}}Kp(0.5)/2}$ \cite{aronson80,newman80}; this yields the estimate
\begin{multline*}
	D_{\vec{x}}=\frac{2c^2}{Kp}=\Bigl(\frac{6.2-2.6\;\text{mm}}{40\times 24\;\text{h}}\Bigr)^2 \times \frac{2}{10^6\times \text{cells/mm}^3  \times 2.16 \times 10^{-2}\,\text{h}^{-1}} \\
	\approx \frac{1.30\times 10^{-3}\; \text{mm}^2/\text{h}}{10^6\times \text{cells/mm}^3 } \approx 1.30 \times 10^{-9}\;(\text{mm}\times\text{cells}\times\text{h})^{-1}.
\end{multline*}

We assume that this coefficient is the same also for infected cells, as a priori we have no reason to believe that the infection affects cellular movement. 

The death rate of uninfected tumour cells due to oxygen-driven selective pressure  $\eta$ and the epigenetic diffusion coefficient of tumour cells $D_y$ are not easily accessible in the empirical literature, hence they have been adapted from previous mathematical papers about epigenetically structured populations: their values have been set respectively to $1/48\;$h$^{-1}\approx 4.16 \times 10^{-2}\;$h$^{-1}$, which is of the same magnitude of the value used in Ref. \cite{chiari23heterogeneity}, and $5.00\times 10^{-6}\;$h$^{-1}$, as in Ref. \cite{celora21}.

The maximal infection rate of the oncolytic virus $\beta_M$ has been set to $7.00\times 10^{-10}$ mm$^3$/(viral particles$\times$h), as in Ref. \cite{friedman06}; their model does not explicitly take into account hypoxia, so we assume that they consider normoxic conditions. Since we are not aware of any experimental estimate of infection rate under hypoxic conditions, we set $\beta_m$ to one fourth of the value of $\beta_m$. The death rate of infected cells $q_I$ has been taken equal to $1/24\;$h$^{-1}=4.17\times 10^{-2}\;$h$^{-1}$, following Ref. \cite{ganly00}. The clearance rate of the virus has been set to $1/6$, as in Ref. \cite{mok09}. The viral load released by the death of infected cells depends highly on the type of virus and ranges from the value $157\;$viral particles/cells estimated in Ref. \cite{workenhe14} to the value $3500\;$viral particles/cells of Ref. \cite{chen01}; we chose an intermediate value of $\alpha=1000\;$viral particles/cells. It is important to remark that all these values are highly dependent on the exact type of oncolytic virus employed; our choices allow to model significant differences in the effectiveness of oncolytic virotherapy as the oxygen level varies. We remark that the outcome of the therapy is mostly determined by the aggregate value $\beta \alpha q_I /q_v$, hence similar dynamics may also be obtained by different parameter combinations that maintain the ratio unchanged (see also the discussion in Ref. \cite{morselli23}). The spatial diffusion coefficient of viral particles $D_v$ has been set to $3.6\times 10^{-2}\;$mm$^2$/h, as in Ref. \cite{friedman06}.

We consider the oxygen thresholds defined in Ref. \cite{mckeown14}: the oxygen partial pressure (pO$_2$) in arterial blood is $70\;$mmHg and we consider this as the maximal oxygen concentration ($O_{\max}$); the physiological pO$_2$ ranges approximately between $57\;$mmHg and $30.4\;$mmHg, so we consider the higher value as the normoxic threshold $O_M$, keeping in mind that we may observe lower oxygen values also in healthy tissue; the pathological hypoxic pO$_2$ value is $7.6\;$mmHg, which we consider as $O_m$. All these pressure values are converted in volume ratios by multiplying them by the solubility constant $3\cdot 10^{-5}\,$mm$_{O_2}^3$/(mm$^3{_\text{plasma}}\times$mmHg) \cite{pittman11}.

We assume that the oxygen decay is due to the consumption of the healthy cells in the region. According to \cite{wagner11}, cells have an average rate of oxygen utilisation of $9.00\times 10^{-15}\,$mol/(cell$\times$h), corresponding approximately to $2.02\times 10^{-7}\,$mm$_{\text{O}_2}$/(cells$\times$h), but this value may vary several orders of magnitude among different cell types. We therefore assume that a single healthy cell consumes six times this amount of oxygen when the available oxygen level is at $O_{\max}$ and the consumption scales linearly with the oxygen concentration, meaning that the consumption in the case of unitary cell density is given by $O(t,\vec{x})$ multiplied by
\[
\frac{1.21 \times 10^{-6} \, \text{mm}^3_{\text{O}_2}}{\text{cell}\times \text{h}} \ \frac{1}{O_{\max}}=5.60\times 10^{-4}\; \text{mm}^3/(\text{cell}\times \text{h}).
\]
Considering $K$ as the healthy cell density in the absence of a tumour, we obtain a decay rate $q_O=5.60\times 10^{2}\;$h$^{-1}$. We adopt a similar way of reasoning for the consumption by cancer cells, starting from the fact that the consumption of a single cell is estimated to be $2.62\times 10^{-6}\,$mm$_{\text{O}_2}$/(cells$\times$h) \cite{grimes14} and assuming again that this is only possible when the oxygen level is $O_{\max}$. We then assume that cancer cells take the place of healthy cells, meaning that they cause an additional consumption of
\[
\lambda=\frac{2.62\times 10^{-6} \text{mm}_{\text{O}_2}/(\text{cells}\times\text{h})}{O_{\max}} -\frac{q_O}{K} = 6.55 \times 10^{-4}\;\text{mm}^3/(\text{cell}\times \text{h}).
\]

The oxygen diffusion coefficient $D_O$ has been set to $3.60\;$mm$^2$/h, as in Ref. \cite{mueller-klieser84}.

Numerical simulations are run until the final time $T=2500\;$h, since their behaviour up to that moment is also representative of later dynamics. For the spatial domain $\Omega=[-L,L]^2$ , we set $L=10\;$mm so that in most cases wave fronts do not hit the boundary before $T$ and the domain is representative of typical extensions of solid tumours.

\subsection{Numerical method}

Numerical simulations use a finite volume method developed by adapting the procedures presented in Refs. \cite{bessemoulin12,carrillo15} to our problem. We discretise the space $\Omega$ with a uniform mesh consisting of the cells
\[
C_{j,k} \coloneqq [x_{1,j-\frac{1}{2}}, x_{1,j+\frac{1}{2}}] \times [x_{2,k-\frac{1}{2}}, x_{2,k+\frac{1}{2}}]
\]
for ${j=0,\dots, N_{x_1}}$, ${k=0,\dots, N_{x_2}}$.
Similarly, we discretise the space $\Omega \times Y$ with the cells
\[
C_{j,k,m} \coloneqq [x_{1,j-\frac{1}{2}}, x_{1,j+\frac{1}{2}}] \times [x_{2,k-\frac{1}{2}}, x_{2,k+\frac{1}{2}}] \times [y_{m-\frac{1}{2}}, y_{m+\frac{1}{2}}]
\]
for ${j=0,\dots, N_{x_1}}$, ${k=0,\dots, N_{x_2}}$, and  ${m=0,\dots, N_{y}}$. The sizes of the cells are thus respectively $\Delta x_1\times\Delta x_2$ and  $\Delta x_1\times\Delta x_2\times\Delta y$, where 
\[
\Delta x_i= \frac{2L}{N_{x_i}+1} \quad  (i=1,2) \qquad \Delta y= \frac{1}{N_y+1}
\]
We set $N_{x_i}=200$ and $N_y=20$. 

The equations for infected cells, viruses and oxygen are not epigenetically structured and are of the form
\begin{equation*}
	\pt f(t,\vec{x})= \mathcal{M} (t,\vec{x}) + \mathcal{R}(t,\vec{x})
\end{equation*}
where $\mathcal{M} (t,\vec{x})$ regulates the movement and $\mathcal{R}(t,\vec{x})$ the reactions. We adopt a splitting method, considering separately the movement and reaction terms. The quantity of our interest is
\[
f_{j,k}(t)= \frac{1}{\Delta x_1\Delta x_2} \int\limits_{C_{j,k}} f(t,\vec{x}) \di\vec{x}.
\]
We begin with the conservative part $\mathcal{M}(t,x)$, which is given by
\begin{equation*} 
	\mathcal{M} (t,\vec{x}) = D\Delta_{\vec{x}} f = D\dive_{\vec{x}}(\nabla_{\vec{x}}  f)
\end{equation*}
in the case of virus and oxygen and by
\begin{equation*}
	\mathcal{M} (t,\vec{x}) =D \dive_{\vec{x}} (\Phi(t,\vec{x}) f ), \qquad \Phi(t,\vec{x})=\nabla_{\vec{x}}\rho (t,\vec{x})
\end{equation*}
in the case of infected cells. In both situations, $\mathcal{M}$ involves the divergence of some quantity, hence the semi-discrete scheme takes the form
\begin{equation}
	\label{eq:semidiscrete}
	\frac{\di f}{\di t}=-D\frac{M_{j+\frac{1}{2},k}(t) - M_{j-\frac{1}{2},k}(t)}{\Delta x_1} -D\frac{M_{j,k+\frac{1}{2}}(t) - M_{j,k-\frac{1}{2}}(t)}{\Delta x_2}
\end{equation}
where $M$ is given by
\[
M_{j+\frac{1}{2},k} = - \partial_{x_1}f_{j+\frac{1}{2},k} \qquad M_{j,k+\frac{1}{2}} = - \partial_{x_2}f_{j,k+\frac{1}{2}}
\]
in the case of spatial diffusion and by
\begin{gather*}
M_{j+\frac{1}{2},k} = (\Phi^1_{j+\frac{1}{2},k})^{+}f_{j,k} + (\Phi^1_{j+\frac{1}{2},k})^{-}f_{j+1,k} \\
M_{j,k+\frac{1}{2}} = (\Phi^2_{j,k+\frac{1}{2}})^{+}f_{j,k} + (\Phi^2_{j,k+\frac{1}{2}})^{-}f_{j,k+1}
\end{gather*}
in the case of pressure-driven movement. In this second case, $\Phi^1 := \partial_{x_1}\Phi$ and $\Phi^2 := \partial_{x_2}\Phi$ are the components of $\Phi$ along the $x_1$ and $x_2$ axis respectively; furthermore, $(\cdot)^+$ and $(\cdot)^-$ indicate respectively the positive and negative part of their arguments, i.e., ${(\cdot)^+= \max\{0,\cdot\}}$ and ${(\cdot)^-= \min\{0,\cdot\}}$. Since our scheme is of order zero, the reconstruction of the function is piecewise constant and thus assumes the same values at all the interfaces. In all the cases, the derivatives in the middle points are evaluated as
\[
(\partial_{x_1} f)_{{j+\frac{1}{2}},k} = \frac{f_{j+1,k}-f_{j,k}}{\Delta x_1}\,, \qquad {(\partial_{x_2}} f)_{j,k+\frac{1}{2}} = \frac{f_{j,k+1}-f_{j,k}}{\Delta x_2}
\]
while the derivatives in the nodes are evaluated as
\[
(\partial_{x_1} f)_{j,k} = \frac{f_{j+1,k}-f_{j-1,k}}{2\Delta x_1}, \qquad
(\partial_{x_2} f)_{j,k} = \frac{f_{j,k+1}-f_{j,k-1}}{2\Delta x_2}.
\]

We use a uniform time discretisation of size $\Delta t=5\times 10^{-1}\;$h and denote with apex $l$ the discretised time step, i.e., $t^l = l \Delta t$. At all the iterations, we check that $\Delta t$ satisfies the positivity-preserving CFL for infected cells, namely
\[
\Delta t \leq \Delta \mathcal{T}_l:=\min \left\{\dfrac{\Delta x_1}{4\Phi^1_M},\dfrac{\Delta x_2}{4\Phi^2_M} \right\}
\]
where $\Phi^1_M=\underset{j,k}{\max}\left(\abs{\Phi^{1\; l}_{j+\frac{1}{2},k}}\right)$
and $\Phi^2_M=\underset{j,k}{\max}\left(\abs{\Phi^{2\; l}_{j,k+\frac{1}{2}}}\right)$. In the case of standard diffusion, the CFL is time-independent and is given by 
\[
\Delta t \leq \min \left\{\dfrac{(\Delta x_1)^2}{4D}, \dfrac{(\Delta x_2)^2}{4D} \right\}.
\]
The fast oxygen dynamics require a refined temporal discretisation. Hence, we set $\Delta t_O=6.95\times 10^{-5}\;$ (which is one-tenth of the maximum size required by the CFL condition for oxygen) and perform several sub-iterations just for the oxygen while maintaining all the other quantities constant.

For the reaction term $\mathcal{R}(t,\vec{x})$, we adopt a simple forward Euler method for the time derivative.
We set the discretised initial condition $f_{j,k}^0$ provided for each $j=0,\dots, N_{x_1}$ and for $k=0,\dots,N_{x_2}$, being $f_{j,k}^l$ the numerical approximation of $ f_{j,k}(t^l)$.
The complete splitted numerical scheme reads
\begin{equation*}
	\begin{cases}
		f^{l+\frac{1}{2}}_{j,k} =
		f^{l}_{j,k}-D\dfrac{\Delta t}{\Delta x_1} \left(M^{l}_{j+\frac{1}{2},k} - M^{l}_{j-\frac{1}{2},k}\right) - D\dfrac{\Delta t}{\Delta x_2} \left(M^{l}_{j,k+\frac{1}{2}} - M^{l}_{j,k-\frac{1}{2}}\right) \\[0.5cm]
		f^{l+1}_{j,k}  = f^{l+\frac{1}{2}}_{j,k} + \Delta t \, R^{l+\frac{1}{2}}_{j,k}
	\end{cases}
\end{equation*}
for $l=1,\dots,N_l$. We also set no flux boundary conditions.

The equation for uninfected cells is epigenetically structured and takes the form
\begin{equation*}
	\dfrac{\partial f}{\partial t}(t,\vec{x},y)= \mathcal{M} (t,\vec{x},y) + \mathcal{R}(t,\vec{x},y)
\end{equation*}
where 
\begin{equation*} 
	\mathcal{M} (t,\vec{x},y) = D_{\vec{x} }\dive_{\vec{x}} (\Phi(t,\vec{x}) f ) + D_y\Delta_{y} f.
\end{equation*}
Therefore, we are now interested in the quantity
\[
f_{j,k}(t)= \frac{1}{\Delta x_1\Delta x_2\Delta_y} \int\limits_{C_{j,k,m}} f(t,\vec{x},y) \di\vec{x}\di y.
\]
The semi-discrete scheme of the equation takes a form analogous to Eq. \eqref{eq:semidiscrete}, which requires the following definitions:
\begin{gather*}
	M_{j+\frac{1}{2},k,m} = (\Phi^1_{j+\frac{1}{2},k,m})^{+}f_{j,k,m} + (\Phi^1_{j+\frac{1}{2},k,m})^{-}f_{j+1,k,m}, \\
	M_{j,k+\frac{1}{2},m} = (\Phi^2_{j,k+\frac{1}{2},m})^{+}f_{j,k,m} + (\Phi^2_{j,k+\frac{1}{2},m})^{-}f_{j,k+1,m}, \\
	N_{j,k,m+\frac{1}{2}} = -\partial_{y}f_{j,k,m+\frac{1}{2}}.
\end{gather*}
The corresponding splitted numerical scheme is
\begin{equation*}
	\begin{cases}
		f^{l+\frac{1}{3}}_{j,k,m} =
		f^{l}_{j,k,m}-\dfrac{\Delta t}{\Delta x_1} \left(M^{l}_{j+\frac{1}{2},k,m} - M^{l}_{j-\frac{1}{2},k,m}\right) - \dfrac{\Delta t}{\Delta x_2} \left(M^{l}_{j,k+\frac{1}{2},m} - M^{l}_{j,k-\frac{1}{2},m}\right) \\[8pt]
		f^{l+\frac{2}{3}}_{j,k,m}  = f^{l+\frac{1}{3}}_{j,k,m} -\dfrac{\Delta t}{\Delta y} \left(N^{l+\frac{1}{3}}_{j,k,m+\frac{1}{2}} - N^{l+\frac{1}{3}}_{j,k,m-\frac{1}{2}}\right) \\[8pt]
		f^{l+1}_{j,k,m} =	f^{l+\frac{2}{3}}_{j,k,m}+\Delta t\, R^{l+\frac{2}{3}}_{j,k,m}
	\end{cases}
\end{equation*}

\bibliographystyle{siam}
\bibliography{biblio_rev}

\end{document}